\documentclass[aps,prb,,twocolumn,showpacs,floatfix,superscriptaddress]{revtex4-1}
\usepackage[dvips]{graphicx}
\usepackage{dcolumn}
\usepackage{amsmath}
\usepackage{amssymb}
\usepackage{multirow}
\usepackage{color}
\usepackage{moreverb}
\usepackage{bm}
\usepackage{tabularx}
\usepackage[version=3]{mhchem} 
\usepackage[english]{babel}
\usepackage[utf8]{inputenc}
\usepackage{amsmath}
\usepackage{physics}

\def \etal{{\it et al.~}}

\begin{document}
\title {A systematic study of structural, electronic and optical properties of atomic scale defects in 2D transition metal dichalcogenides MX$_2$ (M = Mo,W; X = S, Se, Te)} 

\author{Soumyajyoti Haldar}
\email{Soumyajyoti.Haldar@physics.uu.se}
\affiliation {Division of Materials Theory, Department of Physics and Astronomy, Uppsala University, Box-516, SE 75120, Sweden}

\author{Hakkim Vovusha}
\affiliation {Division of Materials Theory, Department of Physics and Astronomy, Uppsala University, Box-516, SE 75120, Sweden}

\author{Manoj Kumar Yadav}
\altaffiliation[Current address: ]{Nepal Academy of Science and Technology, Khumaltar, Lalitpur, Nepal.}
\affiliation {Division of Materials Theory, Department of Physics and Astronomy, Uppsala University, Box-516, SE 75120, Sweden}

\author{Olle Eriksson}
\affiliation {Division of Materials Theory, Department of Physics and Astronomy, Uppsala University, Box-516, SE 75120, Sweden}

\author{Biplab Sanyal}
\email[Corresponding author: ]{Biplab.Sanyal@physics.uu.se}
\affiliation {Division of Materials Theory, Department of Physics and Astronomy, Uppsala University, Box-516, SE 75120, Sweden}

\date{\today}

\begin{abstract} 
In this work, we have systematically studied structural, electronic and magnetic properties of atomic scale defects in 2D transition metal dichalcogenides \ce{MX2}, (M = Mo and W; X = S, Se and Te) by density functional theory. Various types of defects, e.g., X vacancy, X interstitial, M vacancy, M interstitial, MX and XX double vacancies have been considered. It has been found that the X interstitial has the lowest formation energy ($\sim$ 1 eV) for all the systems in the X--rich condition whereas for M--rich condition, X vacancy has the lowest formation energy except for \ce{MTe2} systems. Both these defects have very high equilibrium defect concentrations at growth temperatures (1000K-1200K) reported in literature. A pair of defects, e.g., two X vacancies or one M and one X vacancies tend to occupy the nearest possible distance.  No trace of magnetism has been found for any one of the defects considered. Apart from X interstitial, all other defects have defect states appearing in the band gap, which can greatly affect the electronic and optical properties of the pristine systems. Our calculated optical properties show that the defect states cause optical transitions at $\sim$ 1.0 eV, which can be beneficial for light emitting devices. The results of our systematic study are expected to guide the experimental nanoengineering of defects to achieve suitable properties related to band gap modifications and characterization of defect fingerprints via optical absorption measurements.
\end{abstract}

\maketitle

\section{Introduction}
Ever since the first successful fabrication of graphene, a single layered honeycomb structure of carbon atoms by Novoselov \etal in 2004\cite{novo}, there has been a continuous increase in research interests on many other 2D monolayers.
Apart from graphene\cite{ohare} other extensively studied single layered systems are silicene~\cite{okamoto,sugiyama,yang,ohare,rachel}, germanene~\cite{ohare,li,cai,rachel}, stanene\cite{yong,bvan,rachel}, phosphorene~\cite{liu,das} and a number of single layered transition metal dichalcogenides (MX$_2$, M = Transition metal and X = Chalcogen)~\cite{miguel,mak,zeng}. 
These single layered systems, having two dimensional (2D) structures, generally exhibit properties, which are remarkably different from their respective three dimensional bulk phases and thus these systems have become important subjects for experimental and theoretical studies.  From the application point of view, 
the 2D semiconducting systems have the minimum possible thickness and wide band gaps, which make them important  for photovoltaics, sensing, biomedicine, solar cells and catalysis.

It is not uncommon to find atomic scale defects in 2D materials, e.g., in graphene, mono and di-vacancies, Stone-Wales defects etc. have been studied quite extensively.
Vacancies are formed during the fabrication of monolayer MoS$_2$ using sonochemical deposition method. It is known that these vacancies have considerable effects on electronic, magnetic and optical properties.
For example MoS$_2$ has been
found to acquire S vacancies during its fabrication. These 
S vacancies are found to be detrimental for n-type conductivity of MoS$_2$ as they create deep trap states for electrons\cite{zhou,noh}.  
Moreover, the concentration of defects depends on the mode of the fabrication
process. For example, the carrier mobility in MoS$_2$ monolayer fabricated by 
chemical vapor deposition method is 0.02 cm$^2$/V-s\cite{lee} whereas in mechanically 
exfoliated monolayer, it can go up to 10 cm$^2$/V-s\cite{radis}.  
Owing to the importance of role of defects,  there are few 
recent theoretical studies of native defects in MoS$_2$\cite{noh,zhou,dang}.
Tongay \etal have investigated the effect of anion vacancy on photoluminescence of MoS$_2$, MoSe$_2$ and WSe$_2$ using experiment and density functional theory calculations.~\cite{tongay2013} Recently Liu and coworkers have studied the influence of anion and metal vacancies on electronic and optical properties of monolayer MoS$_2$.\cite{feng2014} 
Various defects in monolayer \ce{MoS2} have been investigated previously by means of first principles DFT calculations by different groups~\cite{zhou,mahdi,Komsa:2013bp,Komsa:2015df,KC:2014hi, Ataca:2011jg}. Rotational defects have also been studied theoretically and experimentally recently on \ce{MoSe2} and \ce{WSe2} by Lin \etal~\cite{Lin2015}
 
However, to the best of our knowledge, systematic studies on the role of various point defects and double defects on the electronic and optical properties of transition metal dichalcogenides are not available in the literature.

Therefore, in this paper we have systematically investigated the role of defects in modifying the electronic and optical properties of monolayers of transition metal dichalcogenides, MX$_2$ (M = Mo, W and X = S, Se, Te). We present the most probable defects under various growth conditions, the electronic structures of defected materials and the signature of these defects in optical properties. The paper is organized in the following way. First, we present the computational details followed by the results of optimized structures in presence of defects. Then we report formation energies and equilibrium defect concentrations. Finally, we present electronic structures of defected monolayers and their optical signatures.
  

\section{Computational Details}
\label{sec:method}
\subsection{Basic parameters}
All the calculations have been performed with a monolayer MX$_2$ supercell, where M stands for Mo, W and X stands for S, Se, Te. We have considered both M and X defects (in terms of vacancy or interstitial) in all of the above systems.  The supercell is generated by repeating the primitive cell by five times in $a$ and $b$ directions.  A vacuum of 20 {\AA} is included to avoid the interaction between periodic images in the out-of-plane direction.  The calculations have been performed using a plane-wave based density functional code \textsc{vasp}. \cite{vasp}  The generalized gradient approximation of Perdew, Burke and Ernzerhof ~\cite{PBE,PBEerr} has been used for the exchange-correlation potential.  The structures have been optimized using the conjugate gradient method with the forces calculated using the Hellman-Feynman theorem. The energy and the Hellman-Feynman force thresholds have been kept at 10$^{-5}$ eV and 10$^{-2}$ eV/{\AA} respectively. For geometry optimizations, a 3 $\times$ 3 $\times$ 1 Monkhorst-Pack $k$-grid is used. Total energies and electronic structures are calculated with the optimized structures on a 5 $\times$ 5 $\times$ 1  Monkhorst-Pack $k$-grid. It is worth mentioning that some defect states may result in finite magnetic moments. For that purpose, we have considered spin-polarization in all our calculations. However, all the systems relaxed to non magnetic ground states. Spin-orbit coupling was not included.
\begin{table}[htbp]
\caption{Comparison of in-plane lattice constant $a$ (in {\AA})  and band gaps $E^g$ (in eV) for different monolayers of \ce{MX2} as calculated with PBE functionals together with exprimental values. Experimental values of lattice constants have been taken from Refs.~\onlinecite{lattice,lattice2}.  Experimental values of band gaps have been taken from Refs.~\onlinecite{mos2-gap2,mos2-gap,Zhang2014,Ugeda2014,Ruppert2014,Wang2015,desai2014}.\\}
\label{tab:basic-properties}
\begin{tabular}{cc|ccc|ccc}
\toprule
 & & & & & & & \\
 & & MoS$_2$ & MoSe$_2$ & MoTe$_2$ & WS$_2$ & WSe$_2$ & WTe$_2$\\ \colrule
 & & & & & & & \\
$a$ (\AA )& PBE 	& 3.18 & 3.32 & 3.56 & 3.18 & 3.32 & 3.56 \\
    & Expt. & 3.16 & 3.30 & 3.52 & 3.15 & 3.28 & -- \\ \colrule 
 & & & & & & & \\
$E_g (eV) $ & PBE	& 1.73 & 1.47 & 1.08 & 1.89 & 1.58 & 1.09 \\
& Expt. & 1.95 & 1.58 & 1.1 & 1.99 & 1.65 & --\\
\botrule
\end{tabular}	
\end{table}

Using the above parameters we have computed the in-plane lattice constants and band gaps of different pristine ~\ce{MX2} systems. These computed values along with the experimental values have been tabulated in table~\ref{tab:basic-properties}. Our calculated lattice constants and band gaps are in good agreement with the previous theoretical calculations~\cite{Wang:2012fa, KC:2014hi,Komsa:2015df,Rama:2012eb,Kang:2013ez}

\subsection{Defect formation energy}
The defect formation energy $E_{f}$ is defined as follows. 
\begin{equation}
\label{eq:defect_for}
 E_{f} = E_{\rm defect} - \big [ E_{\rm pristine} + \sum_{\rm i} \rm n_i \mu_i \big]
\end{equation}
where $E_{\rm defect}$ is the total energy of the MX$_2$ supercell with defects and $E_{\rm pristine}$ is the total energy of the MX$_2$ supercell without any defect. $\rm n_i$ denotes the number of $i$ element (M or X atoms) added to or removed (with a negative sign) from the pristine system to create defects. $\mu_i$ is the chemical potential of the element $i$. 

To compute the chemical potential, we have used the formula $\mu_M + 2 \times \mu_X = \mu_{MX_2}$, where $\mu_M$ and $\mu_X$ are the chemical potentials for M and X respectively. The total energy per formula unit of the pristine monolayer MX$_2$ is denoted as $\mu_{MX_2}$. For M rich environment, $\mu_{M,max} = E_{tot}^{M}$, where $E_{tot}^{M}$ is the total energy of a bcc M metal per atom. Thus, $\mu_{X,min}$ can be computed as $\mu_{X,min} = (\mu_{MX_2} - \mu_{M,max})/2$. For the X rich environment, $\mu_{X,max} = E_{tot}^{X}$, where $E_{tot}^{X}$ is the  energy per atom of the bulk crystal of X. Therefore, $\mu_{M,min}$ is computed as $\mu_{M,min} = \mu_{MX_2} - 2 \times \mu_{X,max} $. 

\subsection{Optical properties}
The optical properties are also calculated using the \textsc{vasp} code. \cite{vasp}. The optical properties of these materials can be described using the complex dielectric constant: $\varepsilon(\omega) = \varepsilon_1(\omega) + \varepsilon_2(\omega)$. The dielectric function has been computed in the momentum representation by obtaining the matrix elements between occupied and unoccupied eigenstates. The imaginary part of the dielectric function can be derived from the following formula:~\cite{optical_kresse}
\begin{eqnarray}
\varepsilon_2(\omega) = \frac{4\pi^2e^2}{\Omega}\lim_{q\to\infty} \frac{1}{q^2} \sum\limits_{c,v,\mathbf{k}}2w_\mathbf{k}\delta(\epsilon_{c\mathbf{k}} - \epsilon_{v\mathbf{k}}-\omega) \nonumber \\
\times \bra{\mu_{c\mathbf{k}+e_\alpha \mathbf{q}}}\ket{\mu_{v\mathbf{k}}} \bra{\mu_{c\mathbf{k}+e_\beta \mathbf{q}}}\ket{\mu_{v\mathbf{k}}}^* ,
\end{eqnarray}
where the indices $c$ and $v$ denote, respectively, the conduction and valence band states. The cell periodic part of the wavefunctions at the given $\mathbf{k}$-point is denoted by $\mu_{c\mathbf{k}}$. The real part of the dielectric function is obtained by using Kramers-Kronig transformation:

\begin{eqnarray}
\varepsilon_1(\omega) = 1 + \frac{2}{\pi}P \int\limits_{0}^{\infty}\frac{\varepsilon_2(\omega')\omega'}{\omega'^2-\omega^2+i\eta} d\omega\prime, 
\end{eqnarray}
where P and $\eta$ denote, respectively,  the principal value and the complex shift.  

\section{Results and Discussions}
\label{sec:result}

\subsection{Geometry \& Energetics}
\label{subsec:geom_energy}
\begin{figure*}[htbp]
\begin{center}
\includegraphics[scale=0.7]{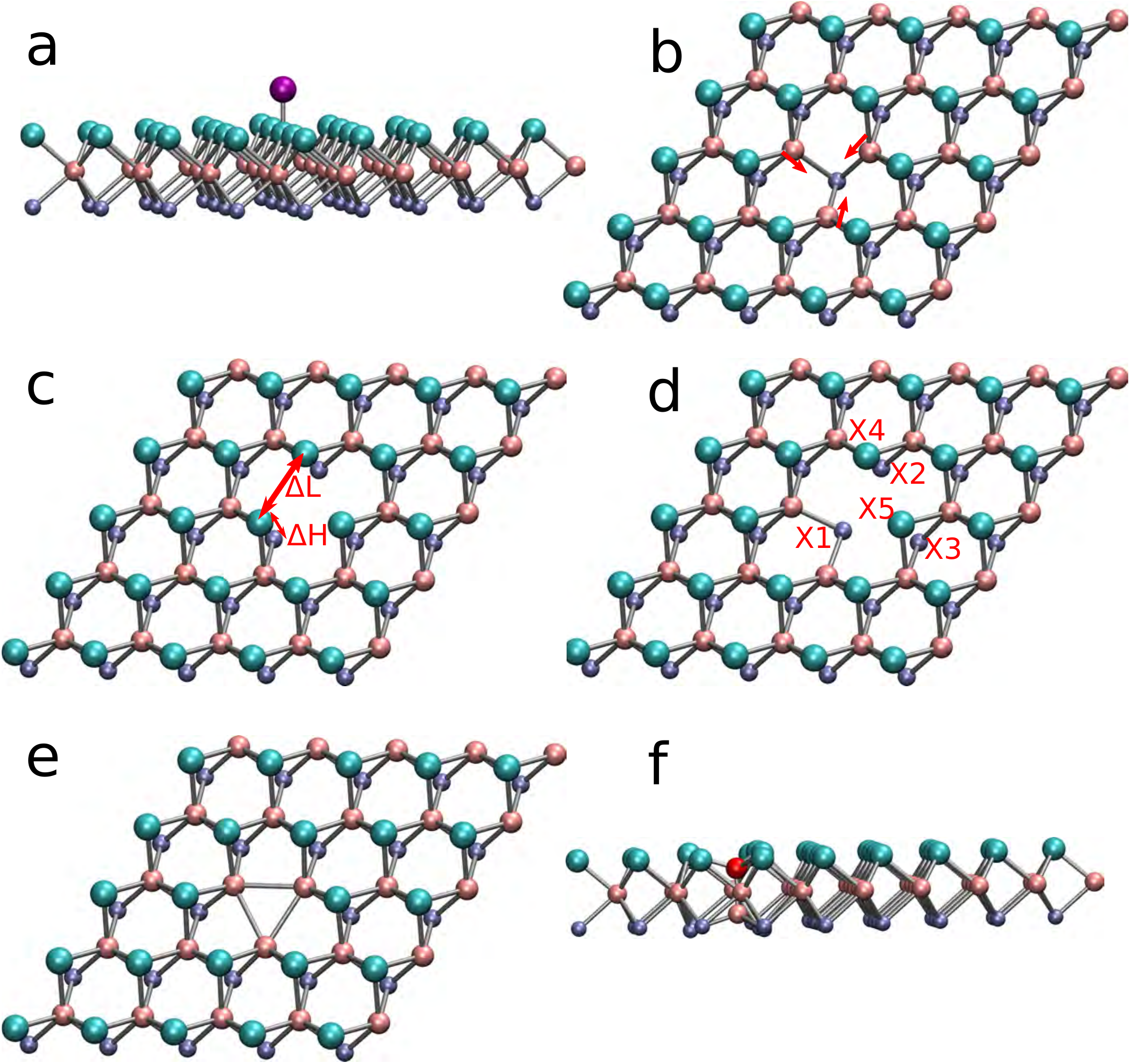}
\caption{(Color online) Representative figure of optimized geometries for various defects in MX$_2$ system as designed on 5$\times$5$\times$1 supercell. Fig.~\ref{fig:rep-geom}(a)-\ref{fig:rep-geom}(f) represent the following defects: (a) X-interstitial, (b) X--vacancy, (c) M--vacancy, (d) MX--vacancy, (e) XX--vacancy and (f) M--interstitial. The cyan (large) balls denote the X--atoms from the top layer, pink (medium) balls denote M--atoms and dark blue (small) balls denote X--atoms from the bottom layer. In Fig.~\ref{fig:rep-geom}(a), the purple ball refers the X interstitial atom. In Fig.~\ref{fig:rep-geom}(f), the red ball indicates to the M interstitial atom. Red arrows in Fig.~\ref{fig:rep-geom}(b) indicates the movement of M atoms during the relaxation. $\Delta L$ and $\Delta H$ in Fig.~\ref{fig:rep-geom}(c) represents the change of in-plane and out-of-plane displacements of the X atoms around the M vacancy due to strutural relaxation. Please see section~\ref{subsec:geom_energy} for detailed analysis of structural relaxations.}
\label{fig:rep-geom}
\end{center}
\end{figure*}
\begin{figure}[htbp]
\begin{center}
\includegraphics[scale=0.45]{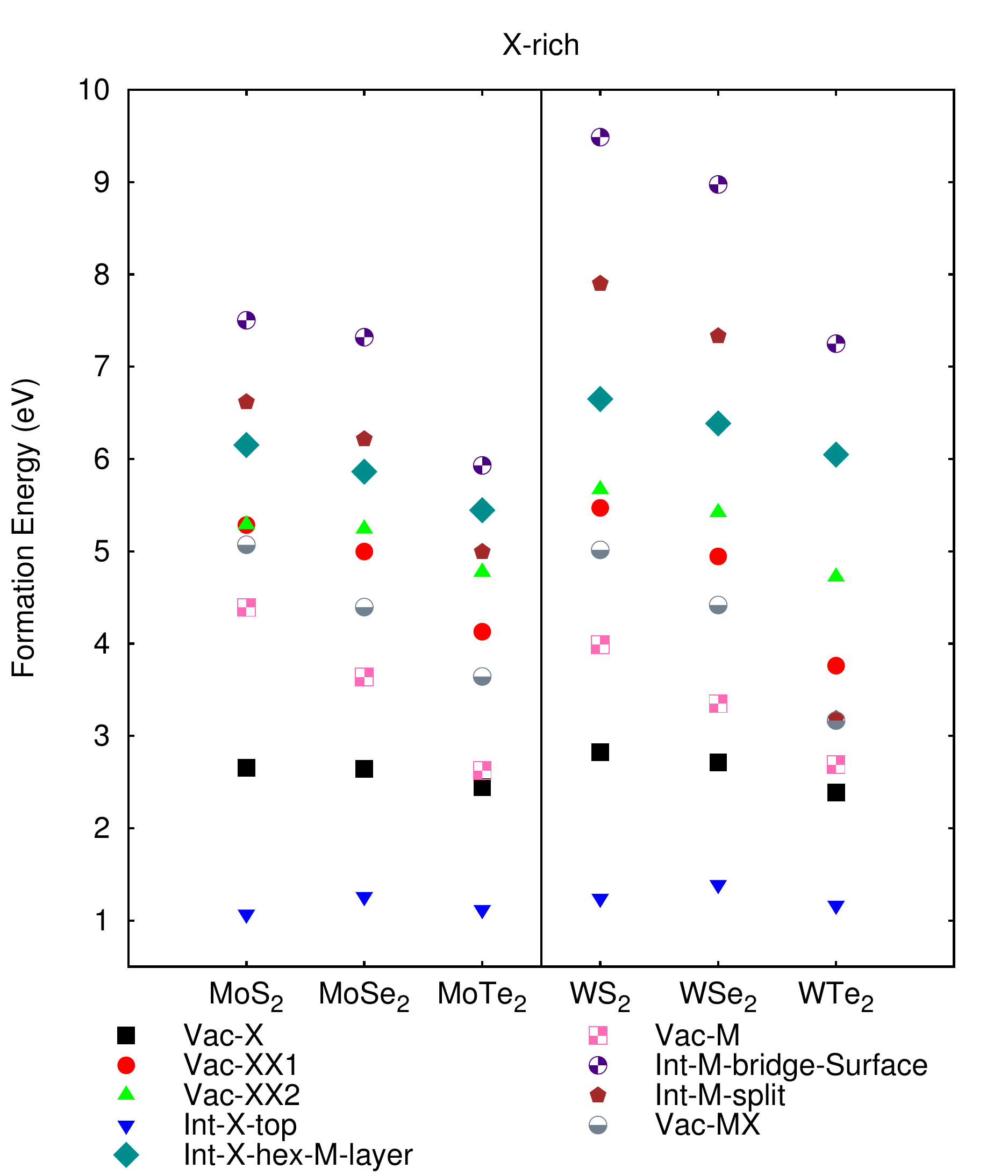}\\
\includegraphics[scale=0.45]{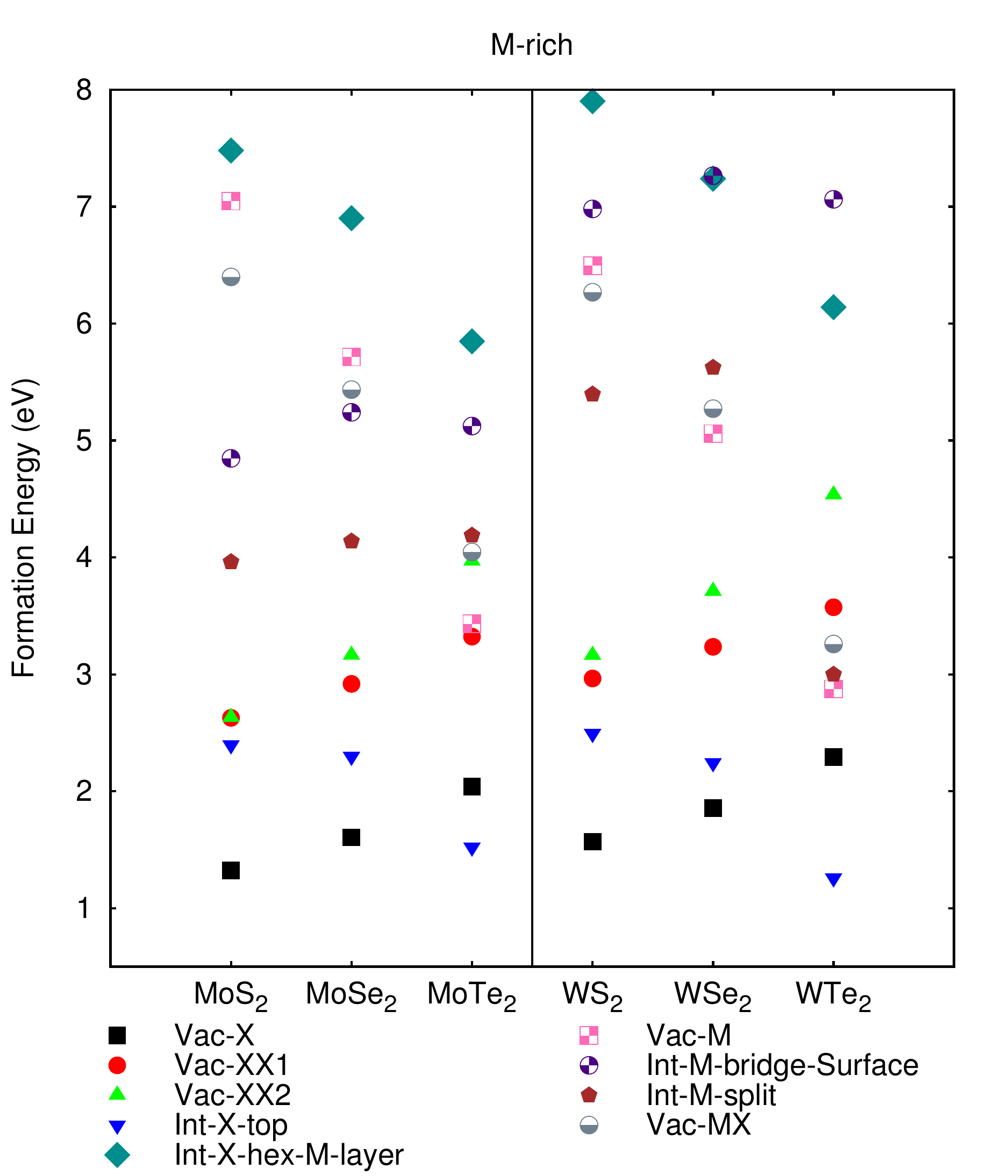}
\caption{(Color online) Formation energies for different types of defects in various MX$_2$ systems under both X--rich and M--rich conditions. See subsection~\ref{subsec:geom_energy} for more details about different systems.}
\label{fig:formation}
\end{center}
\end{figure}
We have compared six different types of defects for various MX$_2$ (M = Mo,W; X = S,Se,Te) systems. The defects are as follows -- X-vacancy, X-interstitial, M-vacancy, M-interstitial, XX-vacancy and MX-vacancy in each monolayer MX$_2$ system. To find out the ground state structures, we have considered many possible starting geometries to find out the lowest energy structure for each of the system. In the following sections, we will first discuss the optimized geometries and energetics of point defects followed by the double defects. In Fig.~\ref{fig:rep-geom} we have shown the typical lowest energy optimized structures of all the different defects that we have considered for the present study. 
  
\subsection*{Point defects}
\label{sec2:single}
\subsubsection*{X -- interstitial}
Fig.~\ref{fig:rep-geom}(a) represents the energetically most stable X interstitial defect (X$_i$) structure. Our calculation of formation energy (See Fig.~\ref{fig:formation}) reveals that under both X--rich and M--rich conditions, the most stable structure is found to have X$_i$ adatom attached to the top of a host X atom (X$_h$). We have considered two other interstitial positions for X$_i$ adatom -- (i) hexagonal position at the M layer and (ii) bridge position between two host X atoms at the top layer. For X$_i$ adatom on top of X$_h$ atom structures, the formation energies are $\sim$ 1.1 eV for all MX$_2$ defects for X--rich environment. However, for M--rich environment, the formation energy drops from $\sim$ 2.4 eV (X = S) to $\sim$ 1.3 eV (for X = Te). The formation energies for the hexagonal position are $\sim$ 6 eV for all MX$_2$ systems, which are quite high compared to the X$_i$ adatom attached to the top of X$_h$ atom. Our calculation also shows that the above mentioned bridge position is a metastable position and after relaxation, the X$_i$ adatom moves to the top of X$_h$ adatom. The X$_i$--X$_h$ bond lengths are 1.94 {\AA}, 2.26 {\AA} and 2.65 {\AA}, respectively for X = S, X = Se and X = Te.

\subsubsection*{X -- vacancy}
The most stable structure of X vacancy is shown in Fig.~\ref{fig:rep-geom}(b). The vacancy is created by removing one X atom in the single layer MX$_2$ supercell. The absence of one X atom causes the three M atoms to relax towards the vacancy site. M--X bond lengths change by 0.04 {\AA} and 0.02 {\AA} respectively for MoX$_2$ and WX$_2$ systems. Under X--rich environment, the formation energy of X vacancy in MoTe$_2$ is 0.21 eV smaller than for MoSe$_2$ and MoS$_2$. The same for WTe$_2$ is 0.44 eV and 0.33 eV smaller than the WSe$_2$ and WS$_2$ respectively. However, formation energy of X vacancy in MTe$_2$ is higher than MSe$_2$ and MS$_2$ under M--rich environment.   

\subsubsection*{M -- vacancy}
\begin{table}[htbp]
\caption{Bond length variation due to the creation of M vacancy with reference to pristine system. $\Delta L$ is the in-plane X--X length variation at the vacancy site and $\Delta H$ is the vertical X--X length variation between the two X atoms from bottom and top layers at the vacancy site. $\Delta_{M-X}$ is the change in the bond length between the X atom with dangling bonds and the M atom.  \\}
\label{tab:bond_mo_vac}
\begin{tabular}{c|ccc|ccc}
\toprule
 & & & & & & \\
Length ({\AA}) & MoS$_2$ & MoSe$_2$ & MoTe$_2$ & WS$_2$ & WSe$_2$ & WTe$_2$\\ \colrule
 & & & & & & \\
$\Delta L$ 	& 0.10	& 0.00	& -0.12	& 0.11 & 0.05 & -0.09 \\
$\Delta H$	& -0.02	& -0.09	& -0.56 & -0.02 & -0.08 & -0.53 \\
$\Delta_{M-X}$	& -0.05	& -0.02	& 0.01 & -0.05 & -0.03 & 0.01 \\ \botrule
\end{tabular}
\end{table}
Fig.~\ref{fig:rep-geom}(c) represents the energetically most stable structure of M vacancy. As the M atom was connected to six X atoms, there are six dangling bonds of X atom present in the structure affecting the relaxation of these X atoms. Table~\ref{tab:bond_mo_vac} refers to the length variation due to creation of the M vacancy. The analysis of length variation shows that while the S and Se atoms relaxed outwards from the vacancy center,  the Te atoms relaxed inwards to the vacancy center. Also, the vertical height between the Te atoms from the top and bottom layer reduces by a large amount ($\sim$ 0.5 {\AA}) compared to the S and Se atoms. The reason behind these geometry changes is that the S and Se are more electronegative compared to Te. Hence the dangling S/Se atoms repel each other strongly leading to the outward relaxation from the vacancy center. For Te atoms, although they repel each other, there is a bigger void to fill in due to the M vacancy. Hence for \ce{MTe2} system, we observe an inward relaxation. These geometry changes directly influence the vacancy formation energies, where MTe$_2$ has the lowest formation energy compared to MS$_2$ and MSe$_2$ under both X--rich and M--rich environment. 

\subsubsection*{M -- interstitial}
\begin{figure}[htbp]
	\begin{center}
		\includegraphics[scale=0.45]{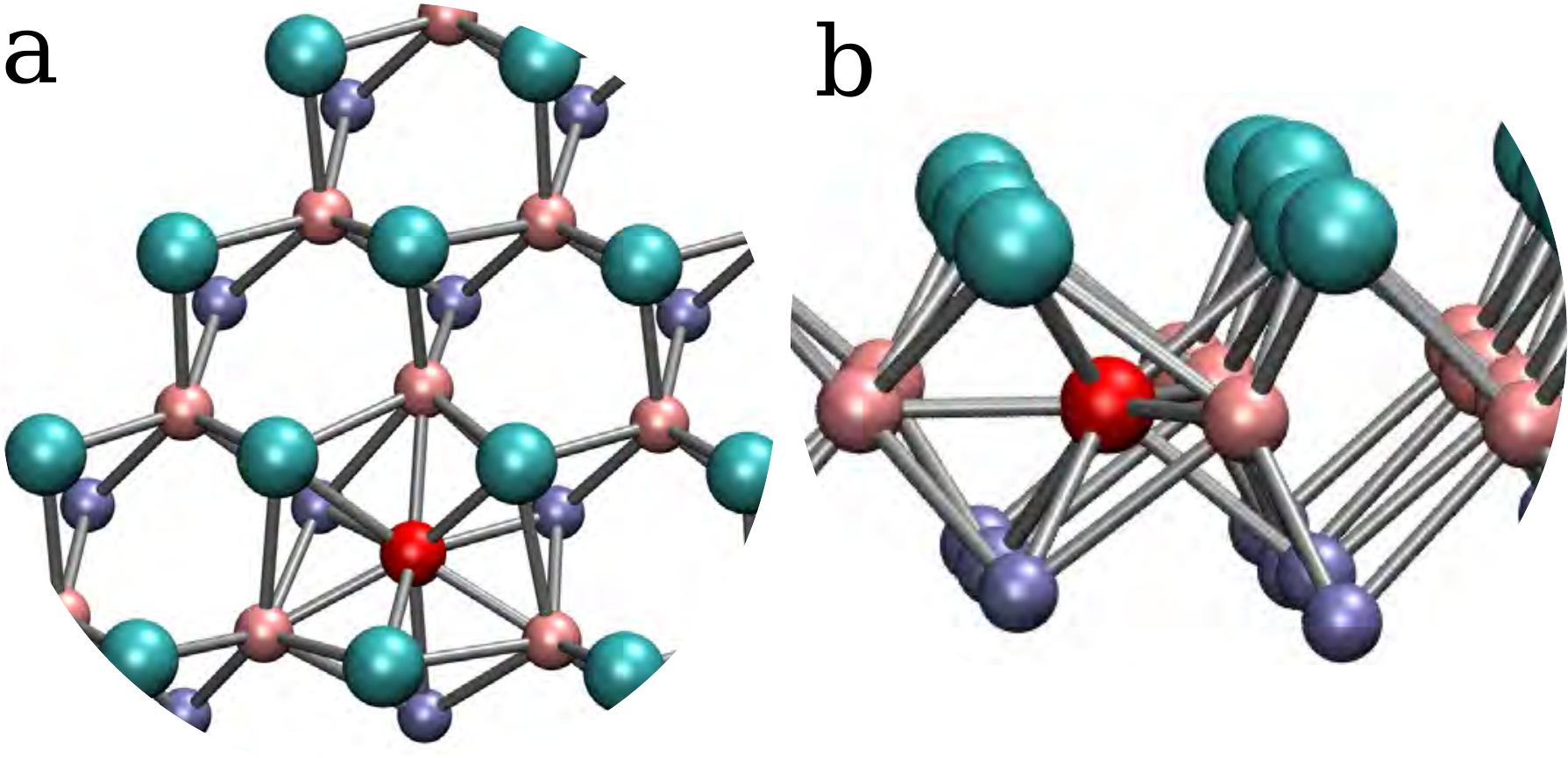}
		\caption{(Color online) Close up view for W interstitial defects in WTe$_2$. (a) slanted top view, (b) side view. The cyan (large) balls denote the X--atoms from the top layer, pink (medium) balls denote M--atoms and dark blue (small) balls denote X--atoms from the bottom layer. The red ball shows the W interstitial atom.
			}
		\label{fig:wte-int-geom}
	\end{center}
\end{figure}
\begin{table}[htbp]
	\caption{M--M$_i$ bond length (in {\AA})variation in M interstitial defect structure in MX$_2$ systems.  \\}
	\label{tab:bond_m_int}
	\begin{tabular}{ccc|ccc}
    \toprule
  & & & & & \\
		 MoS$_2$ & MoSe$_2$ & MoTe$_2$ & WS$_2$ & WSe$_2$ \\ \colrule
  & & & & & \\
		 2.11 & 2.06 & 1.97	& 2.24 & 2.23  \\	 \botrule
	\end{tabular}
\end{table}
Finally, we show the ground state geometry for M interstitial in Fig.~\ref{fig:rep-geom}(f). We have considered two different interstitial positions of M atom. In one case, we have inserted the M atom at the bridge position in between two X atoms and in the other case, the M atom was inserted in the split interstitial position along the $c$ direction. Our calculated formation energy shows that the split interstitial position is energetically most favorable except WTe$_2$. Table~\ref{tab:bond_m_int} shows the bond length variations for M--M$_i$ for the split interstitial position of different \ce{MX2} systems. Fig.~\ref{fig:wte-int-geom} shows the top and side views of energetically favorable structure of W interstitial defect in WTe$_2$, which is quite different from the structure described above. In this case, the interstitial atom resides at the hexagonal position in the M layer forming a distorted hexagon.

\subsection*{Double defects}
\label{sec2:double_defect}
\subsubsection*{MX -- vacancy}
\begin{table}[htbp]
	\caption{Bond length variation due to the creation of a MX vacancy with reference to pristine system. $\Delta L_1$ = change of bond length between X2 and X3 atoms. $\Delta L_2$ = change of bond length between X4 and X5.  $\Delta_{M-X1}$ = change in the bond length between X1 and M atoms.  \\}
	\label{tab:bond_mox_vac}
	\begin{tabular}{c|ccc|ccc}
    \toprule
     & & & & & & \\
		Length ({\AA}) & MoS$_2$ & MoSe$_2$ & MoTe$_2$ & WS$_2$ & WSe$_2$ & WTe$_2$\\ \colrule
     & & & & & & \\
		$\Delta L_1$ 	& 0.22	& 0.08	& -0.35	& 0.23 & 0.16 & 0.08 \\
		$\Delta L_2$	& -0.05	& -0.26	& -0.49 & -0.02 & -0.15 & 0.08 \\
		$\Delta_{M-X1}$	& -0.12	& -0.12	& -0.13 & -0.12 & -0.12 & -0.13 \\ \botrule
	\end{tabular}
\end{table}
\begin{figure}[htbp]
	\begin{center}
		\includegraphics[scale=0.45]{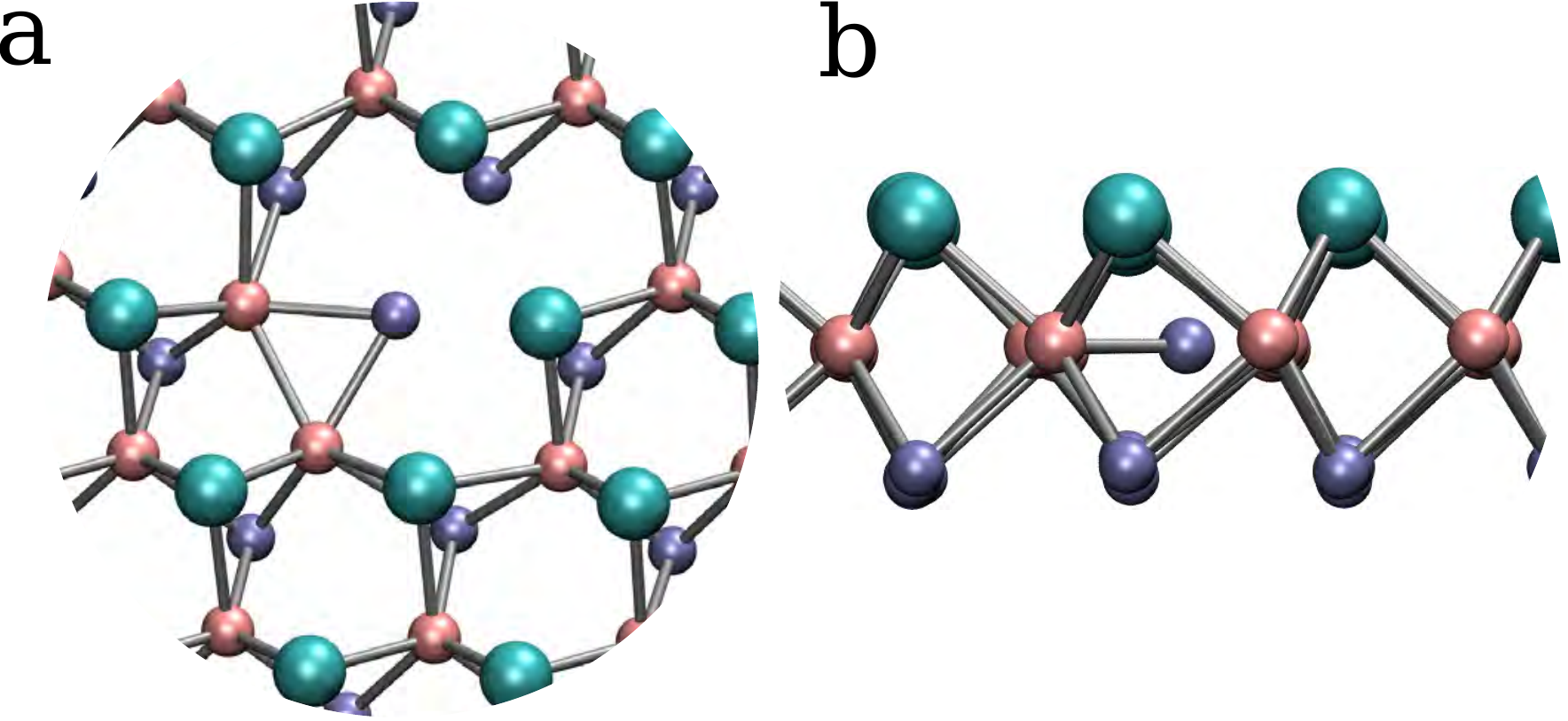}
		\caption{(Color online) Close up view of optimized geometry for WTe vacancy in WTe$_2$. (a) slanted top view, (b) side view. 
			The cyan (large) balls denote the X--atoms from the top layer, pink (medium) balls denote M--atoms and dark blue (small) balls denote X--atoms from the bottom layer.}
		\label{fig:wte-vac-geom}
	\end{center}
\end{figure}
For a MX di-vacancy (created by removing adjacent M and X atoms), the most stable structure is shown in Fig.~\ref{fig:rep-geom}(d). Due to the removal of M and X atoms, five neighboring X atoms around the vacancy site relax during the optimization process. Table~\ref{tab:bond_mox_vac} shows the amount of relaxation of these X atoms with dangling bonds.  After geometry optimization, X1 atom in general moves towards the M layer reducing the M--X1 bond length for all the systems. For MX$_2$ (X = S, Se), the X2 and X3 atoms relax outward from the vacancy site while X4 and X5 atoms relax inward mainly because of the missing X atom in that layer. The reason for these relaxation behavior is similar to the case of M vacancy. For \ce{MoTe2}, the dangling X atom relaxes inward. However, for \ce{WTe2} system they relax slightly outward because of the fact that X1 atom moves to M layer and forms an isosceles triangle with two M atoms with bond length 2.6 {\AA}. Fig.~\ref{fig:wte-vac-geom} shows the slanted top and side views of this particular geometry. The formation energy of MX vacancy is lower for MTe$_2$ for both X and M rich environment.

\subsubsection*{XX -- vacancy}
The most stable structure of XX di-vacancy is shown in Fig.~\ref{fig:rep-geom}(e). We have considered two different configurations for XX vacancy. In one configuration, we have removed two X atoms (with same x and y coordinates) from the top and bottom layers (vac--XX1) and in the other configuration, two nearest X atoms have been removed from the same layer (vac--XX2). The calculation of formation energies indicates that the vac--XX1 structure is energetically more favorable compared to vac--XX2 structure. Here we find that the M atoms relax inward to the vacancy site and form an equilateral triangle with bond length $\sim$ 2.8 \AA. 

\begin{table}[htbp]
	\caption{Relative defect formation energies (in eV) for XX- and MX-vacancies for different distances between vacancies (upper panel of the table). The corresponding distances (in {\AA}) between the vacancies (before relaxation) are also mentioned in the bottom panel of the table, where M stands for Mo and W.   \\}
	\label{tab:ef_dist1}
	\begin{tabular}{c||ccc|cc||cc}
    \toprule
     & & & & & & & \\
	 & XX1$_1$ & XX1$_2$ & XX1$_3$ & XX2$_1$ & XX2$_2$ & MX$_1$ & MX$_2$ \\ \colrule
	 & & & & & & & \\
\ce{MoS2} & 0.0 & 0.13 & 0.04 & 0.00 & 0.05 & 0.0 & 2.02 \\
\ce{MoSe2} & 0.0 & 0.45 & 0.32 & 0.24 & 0.35 & 0.0 & 1.88 \\
\ce{MoTe2} & 0.0 & 0.99 & 0.83 & 0.64 & 0.86 & 0.0 & 1.59 \\
     & & & & & & & \\
\ce{WS2} & 0.0 & 0.31 & 0.18 & 0.2 & 0.21 & 0.0 & 1.89 \\
\ce{WSe2} & 0.0 & 0.68 & 0.50 & 0.47 & 0.54 & 0.0 & 1.95 \\
\ce{WTe2} & 0.0 & 1.32 & 1.03 & 0.96 & 1.08 & 0.0 & 1.84 \\
\botrule
	\end{tabular}
~\\
\begin{tabular}{c||ccc|cc||cc}
    \toprule
     & & & & & & & \\
	 & XX1$_1$ & XX1$_2$ & XX1$_3$ & XX2$_1$ & XX2$_2$ & MX$_1$ & MX$_2$ \\ \colrule
	 & & & & & & & \\
\ce{MS2} & 3.12 & 4.46 & 6.33 & 3.18 & 5.59 & 2.41 & 3.99 \\
\ce{MSe2} & 3.34 & 4.71 & 6.65 & 3.32 & 5.75 & 2.54 & 4.18 \\
\ce{MTe2} & 3.60 & 5.07 & 7.14 & 3.56 & 6.17 & 2.73 & 4.49 \\
\botrule
	\end{tabular}
\end{table} 

To investigate the dependence of distance on the formation energies of XX- and MX-vacancies, we have calculated defect formation energies for various distances between a pair of defects. In table~\ref{tab:ef_dist1}, relative  energies as a function of distances are tabulated. XX1 and XX2 vacancies denote two X atoms removed from different layers and same layers of X atom, respectively, from a pristine \ce{MX2} system. XX1$_1$, XX1$_2$ and XX1$_3$ are three different XX1 configurations considered with three different distances between X atoms where XX1$_1$ corresponds to the nearest neighbor configuration.  In a same manner, XX2$_1$ and XX2$_2$ are two different XX2 configurations considered. For MX vacancy also we have considered two different distances -- MX$_1$ and MX$_2$ where MX$_1$ is the nearest neighbor. Our results clearly show that the double vacancies always prefer to form when they are within the nearest neighbor distance signifying the natural occurrence of correlated vacancies. A clear trend of increasing formation energies while one goes down in the periodic table from S to Te is observed for both Mo and W in case of anion vacancies. The trend is not absolutely clear for M-X vacancies.

\subsection{Equilibrium defect concentration}
We have also calculated equilibrium defect concentrations for all the defects. The equilibrium defect concentration, $C_{eq}$, can be calculated using the occupation probability as shown below, 
\begin{equation}
C_{eq} = Ne^{-E_f/k_BT}
\end{equation}
where $N$ is the concentration of possible defect sites, $E_f$ is the formation energy of the defect and $T$ is the temperature (1000K -- 1200K) during the crystal growth. 
\begin{table*}[htbp]
\centering
\caption{Equilibrium defect concentration, $C_{eq}$, in cm$^{-2}$ for X--rich condition for growth temperature in the range of 1000K--1200K\\}
\label{tab:conc_x}
\begin{tabular}{l|llllll}
\toprule
 & & & & & & \\
  & X$_{int}$ & X$_{vac}$ & M$_{int}$ & M$_{vac}$ & MX$_{vac}$ & XX$_{vac}$ \\ 
\colrule
 & & & & & & \\
\ce{MoS2} & 9.2$\times 10^9$--7.2$\times 10^{10}$ & 96.7--1.6$\times 10^4$ & 5.1$\times 10^{-19}$--1.8$\times 10^{-13}$ & 8.4$\times 10^{-8}$--4.1$\times 10^{-4}$ & 6.3$\times 10^{-11}$--1.1$\times 10^{-6}$ & 2.7$\times 10^{-12}$--7.3$\times 10^{-8}$ \\
\ce{MoSe2} & 1.0$\times 10^9$--1.1$\times 10^{10}$ & 1.1$\times 10^2$--1.8$\times 10^4$ & 5.4$\times 10^{-17}$--8.9$\times 10^{-12}$ & 5.3$\times 10^{-4}$--0.6 & 1.6$\times 10^{-7}$--7.9$\times 10^{-4}$ & 7.5$\times 10^{-11}$--1.2$\times 10^{-6}$\\
\ce{MoTe2} & 5.1$\times 10^9$--4.5$\times 10^{10}$ & 1.1$\times 10^3$--1.2$\times 10^5$ & 7.6$\times 10^{-11}$-- 1.2$\times 10^{-6}$ & 67.1--1.1$\times 10^4$ & 9.9$\times 10^{-4}$--1.14 & 1.8$\times 10^{-6}$--5.2$\times 10^{-3}$\\
\ce{WS2} & 1.2$\times 10^9$--1.4$\times 10^{10}$ & 13.7--3.2$\times 10^3$& 1.7$\times 10^{-25}$--7.5$\times 10^{-19}$ & 8.9$\times 10^{-6}$--1.2$\times 10^{-2}$ & 1.2$\times 10^{-10}$--2.0$\times 10^{-6}$ & 3.1$\times 10^{-11}$--1.2$\times 10^{-8}$\\
\ce{WSe2} & 2.2$\times 10^8$--3.3$\times 10^9$ & 48.6--9.2$\times 10^3$ & 1.3$\times 10^{-22}$--1.8$\times 10^{-16}$ & 1.5$\times 10^{-2}$--9.78 & 1.2$\times 10^{-7}$--6.4$\times 10^{-4}$ & 1.4$\times 10^{-10}$--1.9$\times 10^{-6}$ \\
\ce{WTe2} & 3.0$\times 10^9$--2.9$\times 10^{10}$ & 2.2$\times 10^3$--2.2$\times 10^5$ & 0.1--47.6 & 32.2--5.8$\times 10^3$ & 0.3--1.1$\times 10^2$ & 1.3$\times 10^{-4}$--0.2\\
\botrule
\end{tabular}
\end{table*}
\begin{table*}[htbp]
\caption{Equilibrium defect concentration, $C_{eq}$, in $cm^{-2}$ for M-rich condition for growth temperature in the range of 1000K--1200K\\}
\label{tab:conc_m}
\begin{tabular}{l|llllll}
\toprule
 & & & & & & \\
 & X$_{int}$ & X$_{vac}$ & M$_{int}$ & M$_{vac}$ & MX$_{vac}$ & XX$_{vac}$ \\
\colrule
 & & & & & & \\
\ce{MoS2} & 1.8$\times 10^3$--1.9$\times 10^5$ & 4.8$\times 10^8$--6.2$\times 10^9$ & 1.2$\times 10^{-5}$--2.6$\times 10^{-2}$ & 3.4$\times 10^{-21}$--2.9$\times 10^{-15}$ & 1.3$\times 10^{-17}$--3.0$\times 10^{-12}$ & 64.6 -- 1.0$\times 10^4$\\
\ce{MoSe2} & 5.8$\times 10^3$--5.0$\times 10^5$ & 1.8$\times 10^7$--4.1$\times 10^8$ & 1.6$\times 10^{-6}$--4.7$\times 10^{-3}$ & 1.8$\times 10^{-14}$--1.1$\times 10^{-9}$ & 9.2$\times 10^{-13}$--3.4$\times 10^{-8}$ & 2.2--6.3$\times 10^2$ \\
\ce{MoTe2} & 4.8$\times 10^7$--9.1$\times 10^8$ & 1.2$\times 10^5$--6.0$\times 10^6$ & 8.8$\times 10^{-7}$--2.9$\times 10^{-3}$ & 5.8$\times 10^{-3}$--4.4 & 9.2$\times 10^{-6}$--2.3$\times 10^{-2}$ & 2.0$\times 10^{-2}$-- 12.6 \\
    \ce{WS2} & 6.0$\times 10^2$--7.5$\times 10^4$ & 2.8$\times 10^7$--5.8$\times 10^8$ & 7.3$\times 10^{-13}$--2.5$\times 10^{-8}$ & 2.1$\times 10^{-18}$--6.0$\times 10^{-13}$ & 5.9$\times 10^{-17}$--1.1$\times 10^{-11}$ & 1.3 --4.0$\times 10^2$\\
    \ce{WSe2} & 1.1$\times 10^4$--8.4$\times 10^5$& 9.8$\times 10^5$--3.6$\times 10^7$& 5.2$\times 10^{-14}$--2.7$\times 10^{-9}$ & 3.7$\times 10^{-11}$--6.5$\times 10^{-7}$ & 6.1$\times 10^{-12}$--1.6$\times 10^{-7}$ & 5.6$\times 10^{-2}$--29.3 \\
    \ce{WTe2} & 1.0$\times 10^9$--1.2$\times 10^{10}$ & 6.4$\times 10^3$--5.4$\times 10^5$ & 0.9--2.9$\times 10^2$ & 3.7--9.6$\times 10^2$ & 8.5$\times 10^{-2}$--46.7& 1.1$\times 10^{-3}$--1.1 \\
\botrule
\end{tabular}
\end{table*}
In Tables~\ref{tab:conc_x} and \ref{tab:conc_m}, we have tabulated the range of equilibrium defect concentrations for a range of growth temperatures (1000K -- 1200K) for X--rich and M--rich conditions respectively. This range of temperature is chosen as most of the \ce{MX2} monolayer structures are synthesized experimentally in this range. \cite{cvd-mos2,cvd-mos2-2,cvd-mose2,cvd-mote2,cvd-ws2,cvd-wse2}. Our calculation clearly shows that the presence of $X_{int}$ defect under the X--rich condition and $X_{vac}$ defect under the M--rich condition during the crystal growth is very much likely. However the probability of forming $M_{int}$ defect under X--rich condition and the formation of $M_{vac}$ defect under M--rich condition is very low.

\subsection{Electronic properties}
\label{subsec:elec}
\begin{figure*}[htbp]
	\begin{center}
		\includegraphics[scale=0.55]{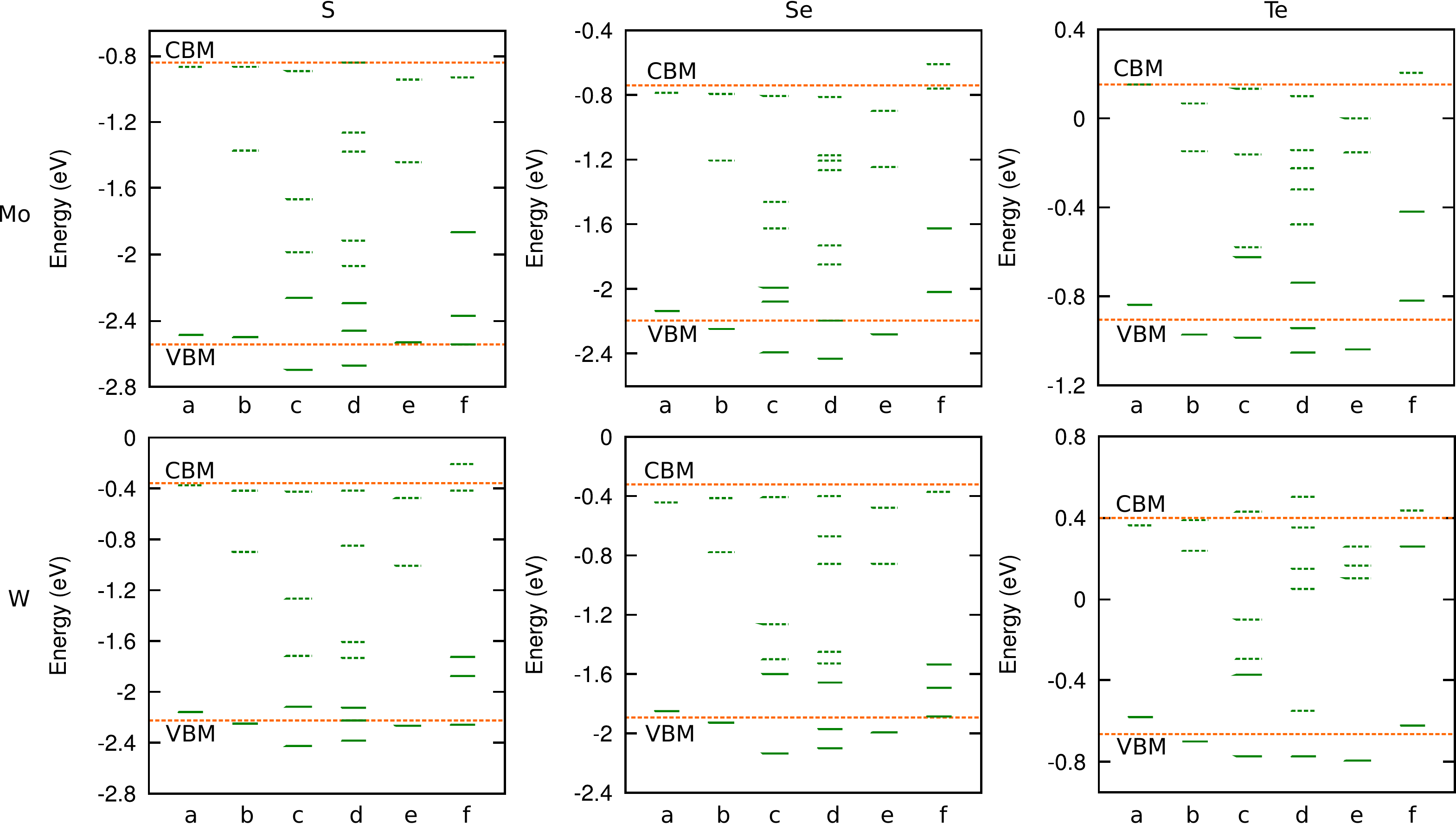}
		\caption{(Color online) Energy level diagram for different defects in all the \ce{MX2} systems. In each subplot, colums (a)-(f) refer to X--interstitial, X--vacancy, M--vacancy, MX--vacancy, XX--vacancy and M--interstitial respectively. The orange (long dashed in print)  line shows the position of VBM and CBM for pristine \ce{MX2}. Green solid and dashed lines denote the occupied and unoccupied defect states respectively.}
		\label{fig:dos-diag}
	\end{center}
\end{figure*}
\begin{figure}[!htp]
	\begin{center}
		\includegraphics[scale=0.35]{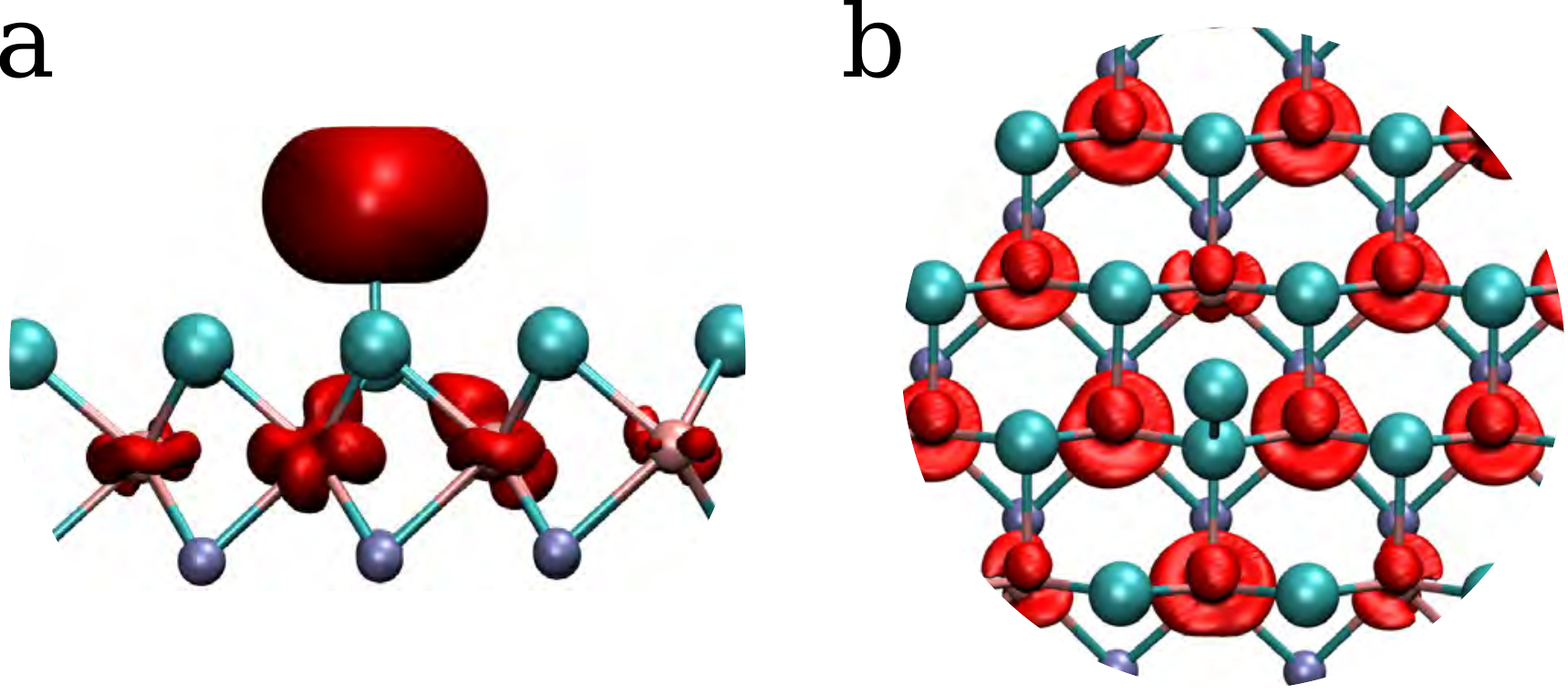}\\
		\includegraphics[scale=0.31]{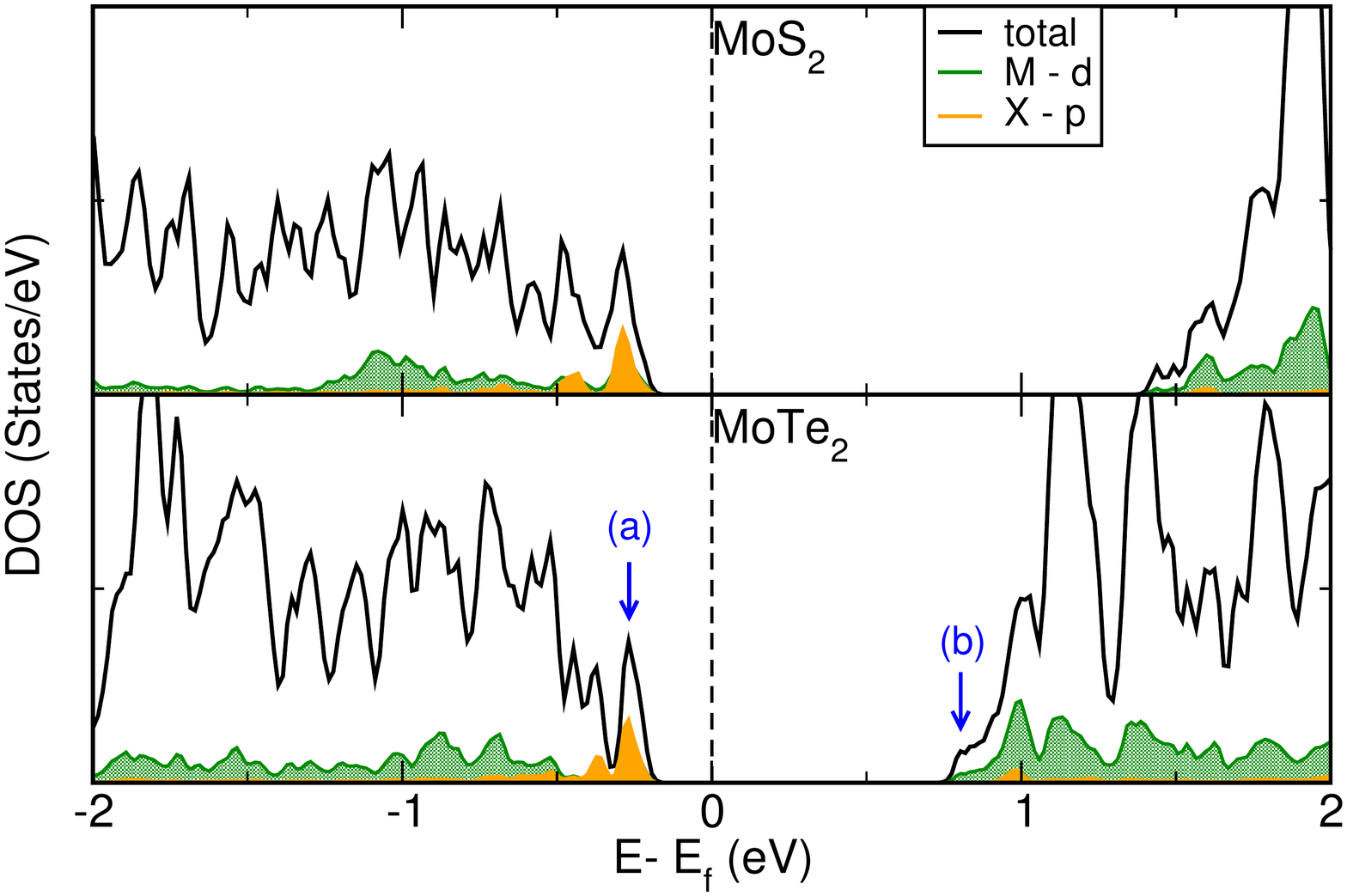}
		\caption{(Color online) Total and partial densities of states for X interstitial in MoS$_2$ and MoTe$_2$. Fig.~\ref{fig:x-int}(a) shows partial charge density of MoTe$_2$ showing electronic orbitals for VBM. Fig.~\ref{fig:x-int}(b) shows the same for CBM. Black (solid) line denotes the total density of states. Green (checker shaded) region represents the defect states due to Mo--$d$ orbitals. Orange (solid shaded) region shows the defect states due to S,Te--$p$ states.}
		\label{fig:x-int}
	\end{center}
\end{figure}
\begin{figure}[!hbp]
	\begin{center}
		\includegraphics[scale=0.35]{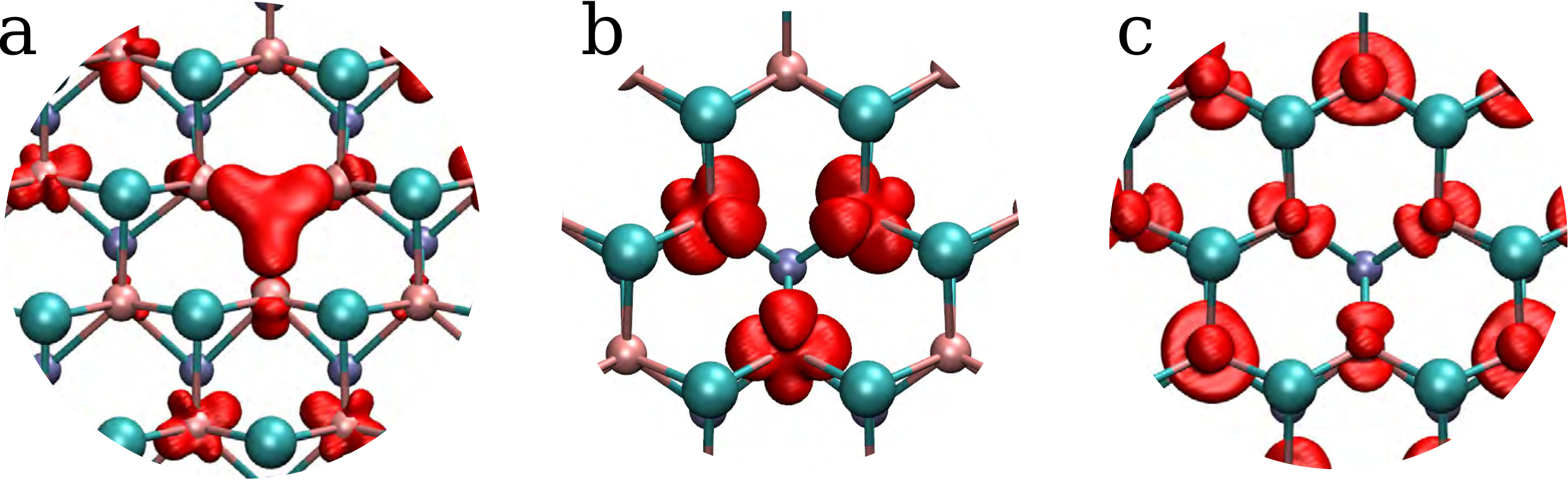}\\
		\includegraphics[scale=0.31]{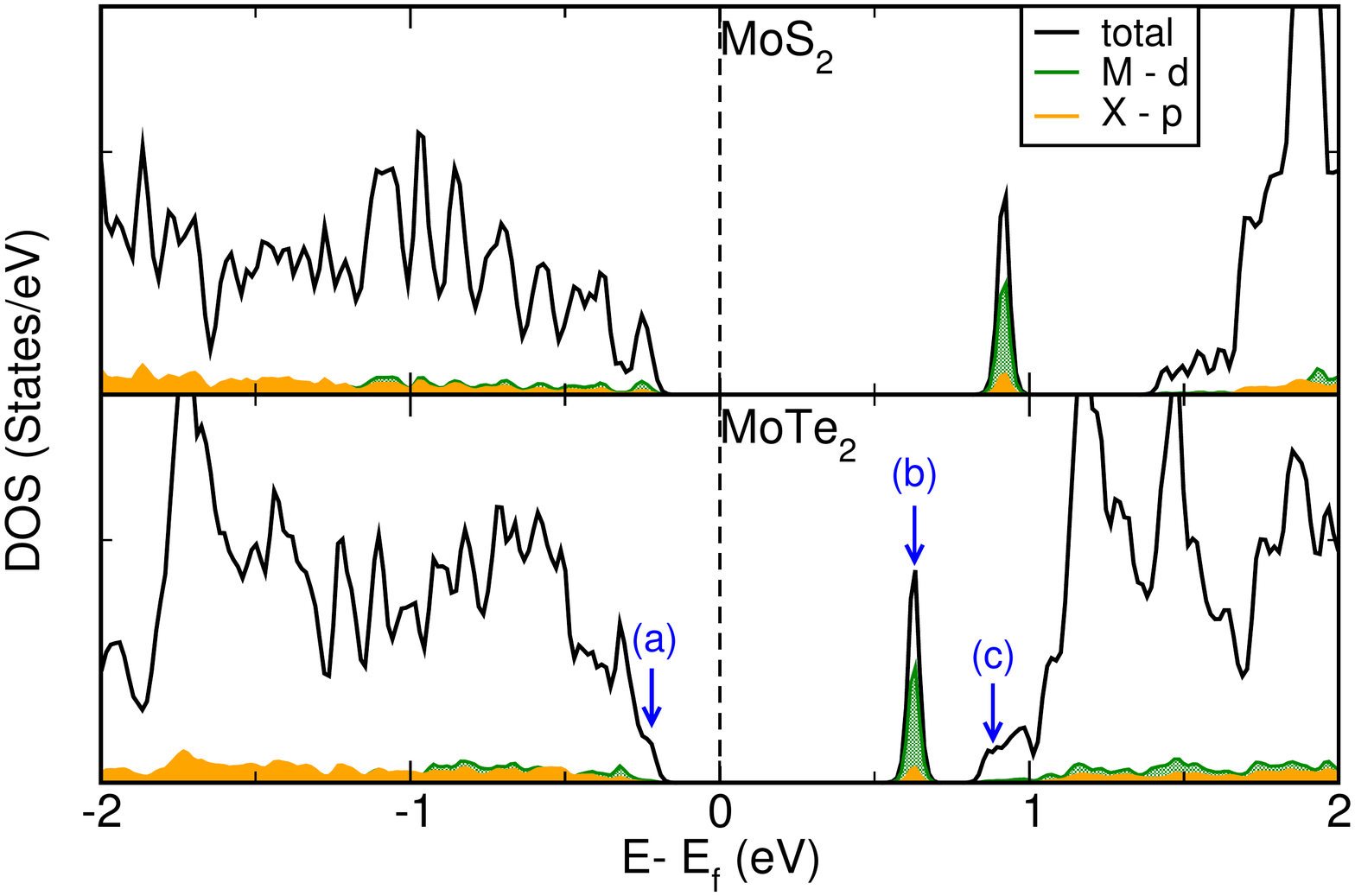}
		\caption{(Color online) Total and partial DOS for X vacancy in MoS$_2$ and MoTe$_2$. Fig.~\ref{fig:x-vac}(a)--(c) show partial charge densities of different defect peaks in the band gap region as indicated in the DOS. See Fig.~\ref{fig:x-int} for other notations.}
		\label{fig:x-vac}
	\end{center}
\end{figure}
\begin{figure}[!hbp]
	\begin{center}
		\includegraphics[scale=0.28]{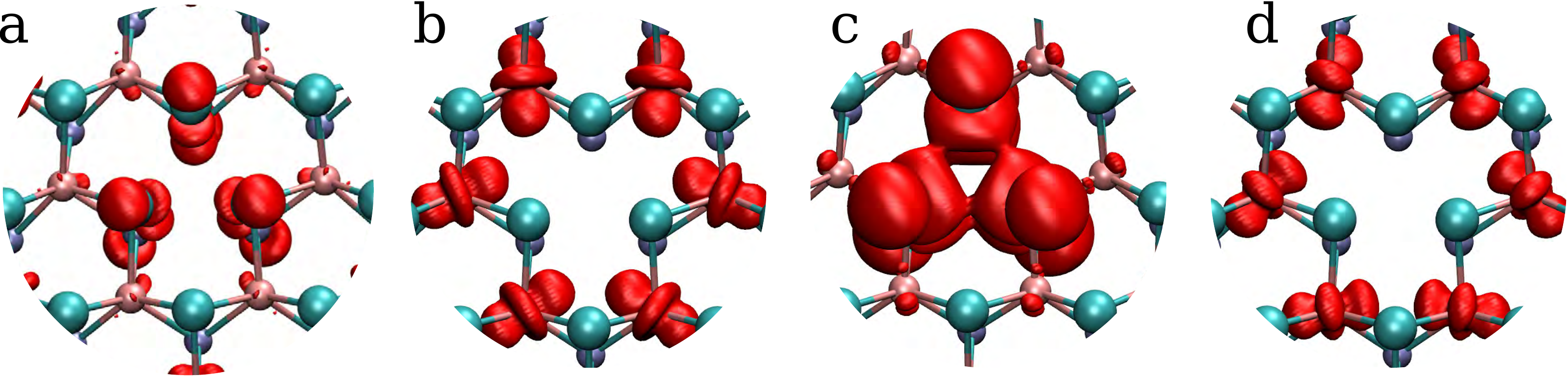}\\
		\includegraphics[scale=0.31]{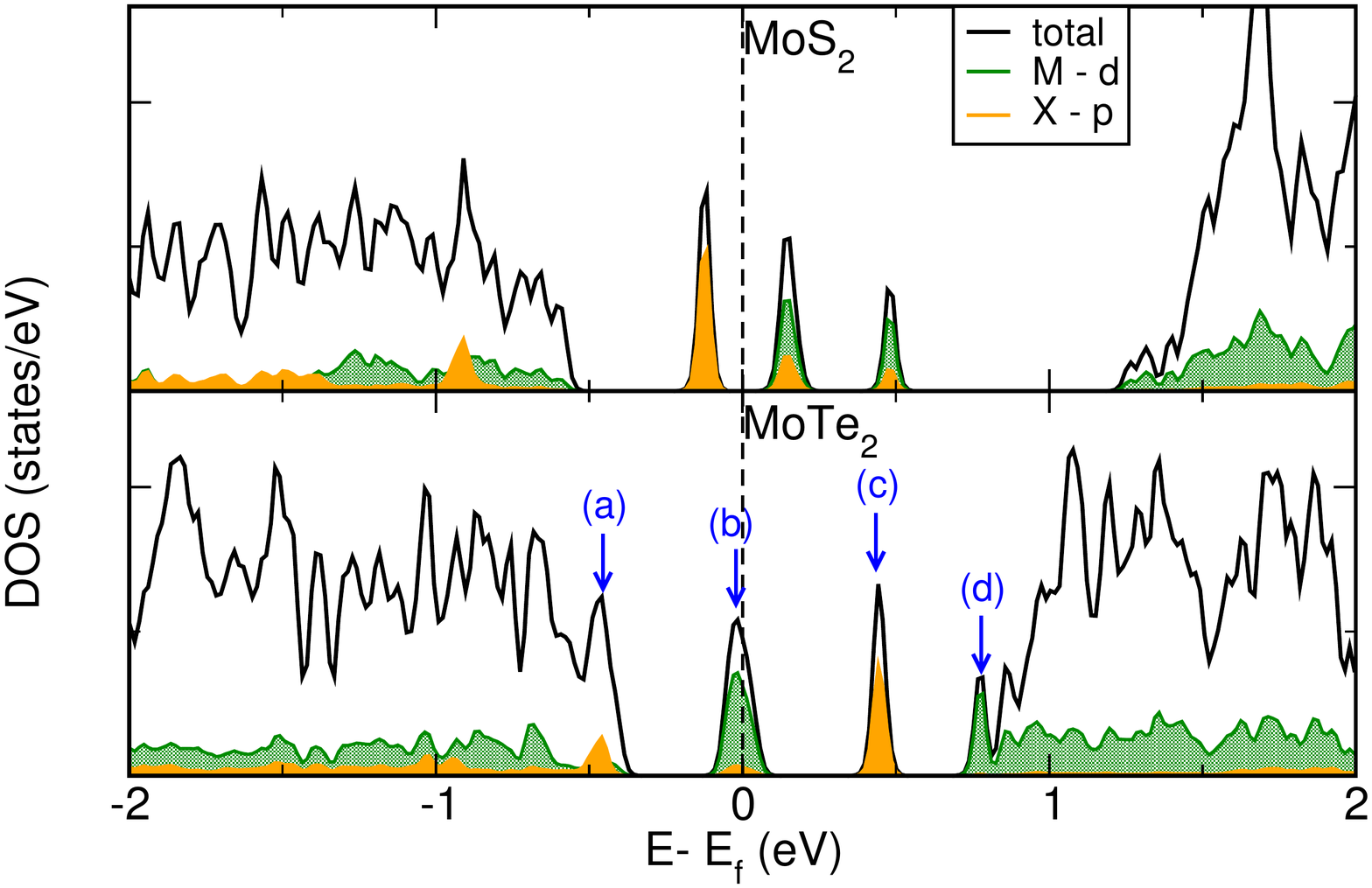}
		\caption{(Color online) Total and projected DOS for M vacancy in MoS$_2$ and MoTe$_2$. Fig.~\ref{fig:vac-m}(a)--(d) show partial charge densities for defect peaks in the band gap region as noted in the DOS. See Fig.~\ref{fig:x-int} for other notations.}
		\label{fig:vac-m}
	\end{center}
\end{figure}
\begin{figure}[!htbp]
	\begin{center}
		\includegraphics[scale=0.28]{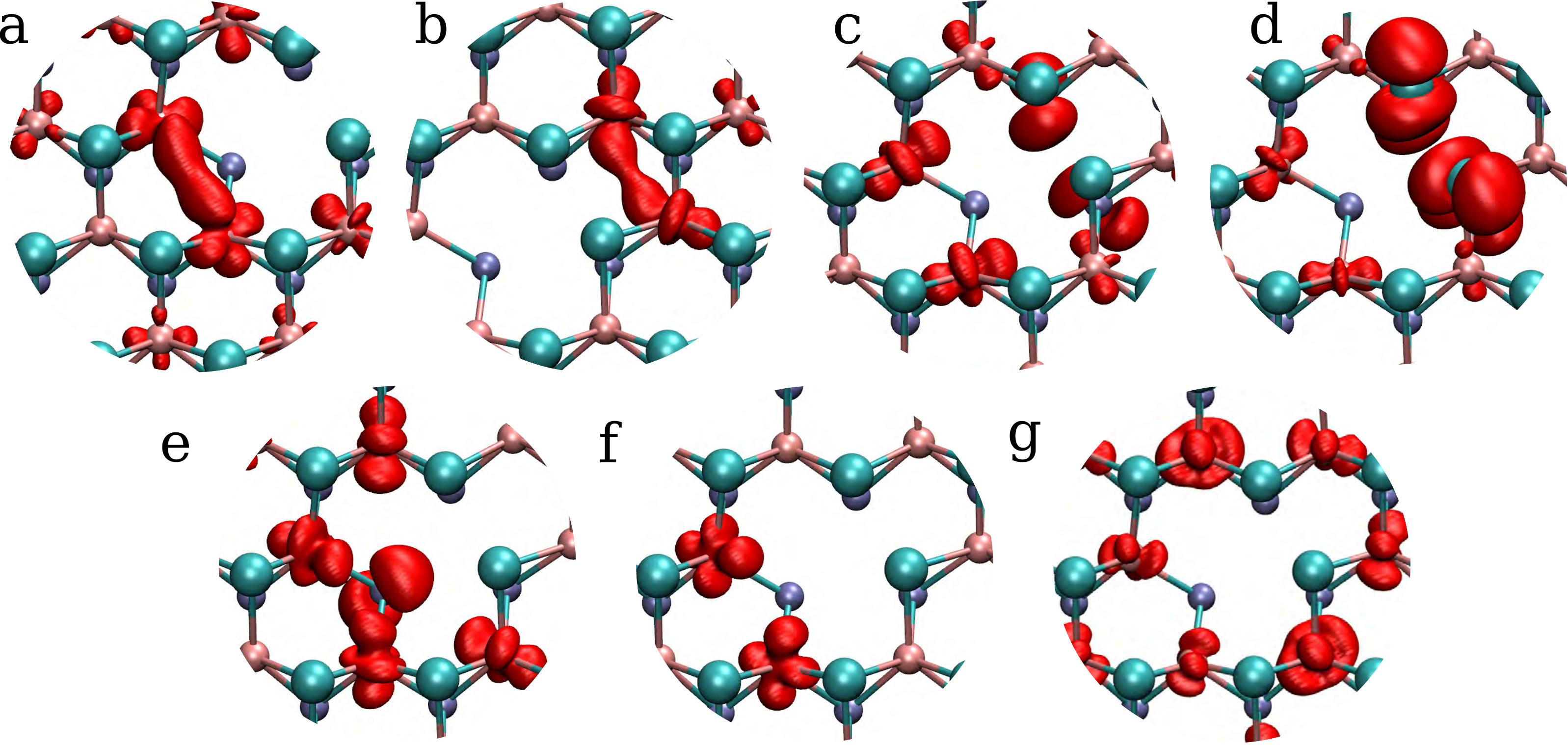}\\
		\includegraphics[scale=0.31]{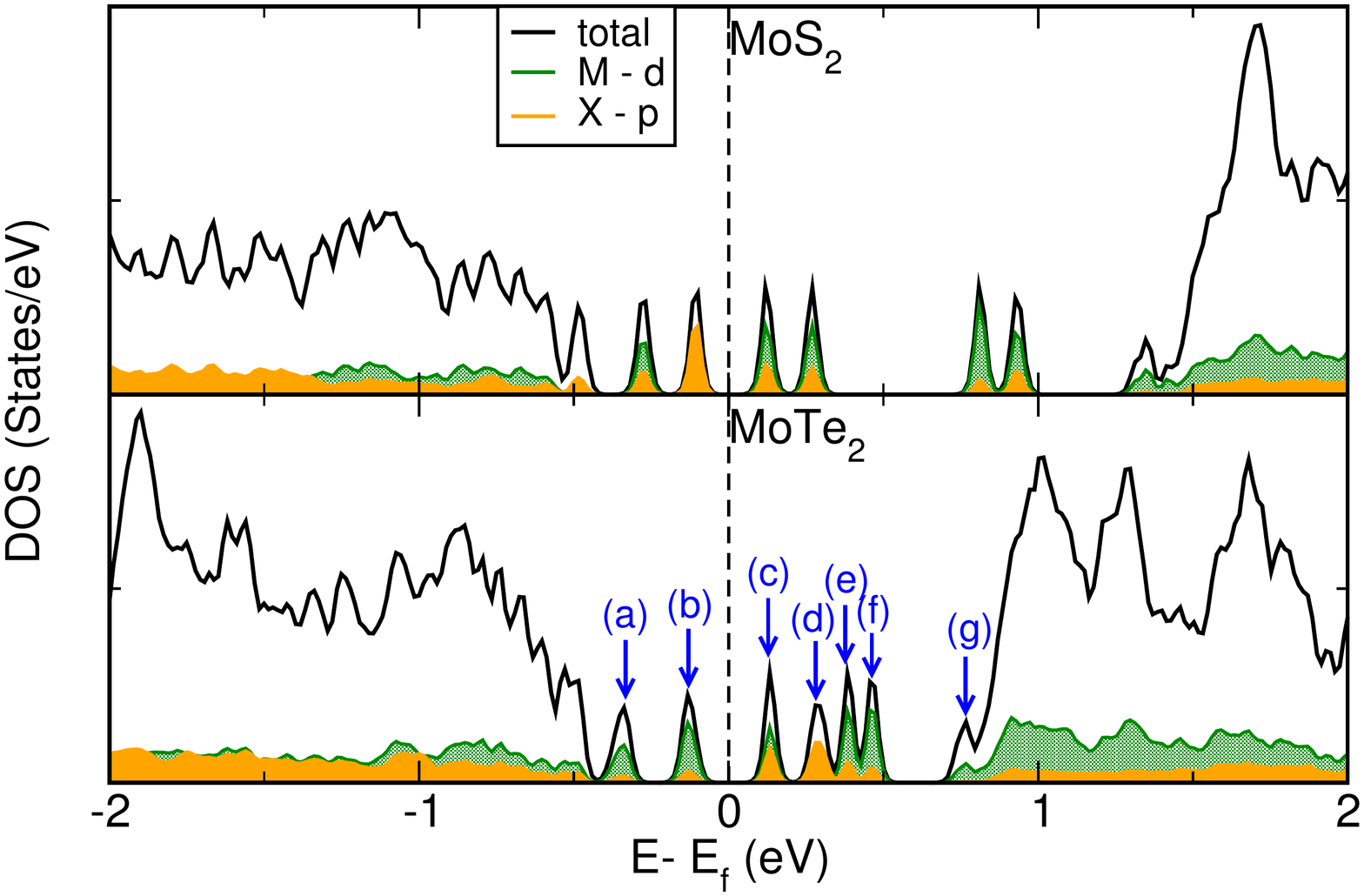}
		\caption{(Color online) Total and partial densities of states for MX vacancy in MoS$_2$ and MoTe$_2$. Fig.~\ref{fig:mx-vac}(a)--(g) show partial charge densities for defect peaks in the band gap region as noted in the DOS. See Fig.~\ref{fig:x-int} for other notations.}
		\label{fig:mx-vac}
	\end{center}
\end{figure}
\begin{figure}[!htbp]
	\begin{center}
		\includegraphics[scale=0.32]{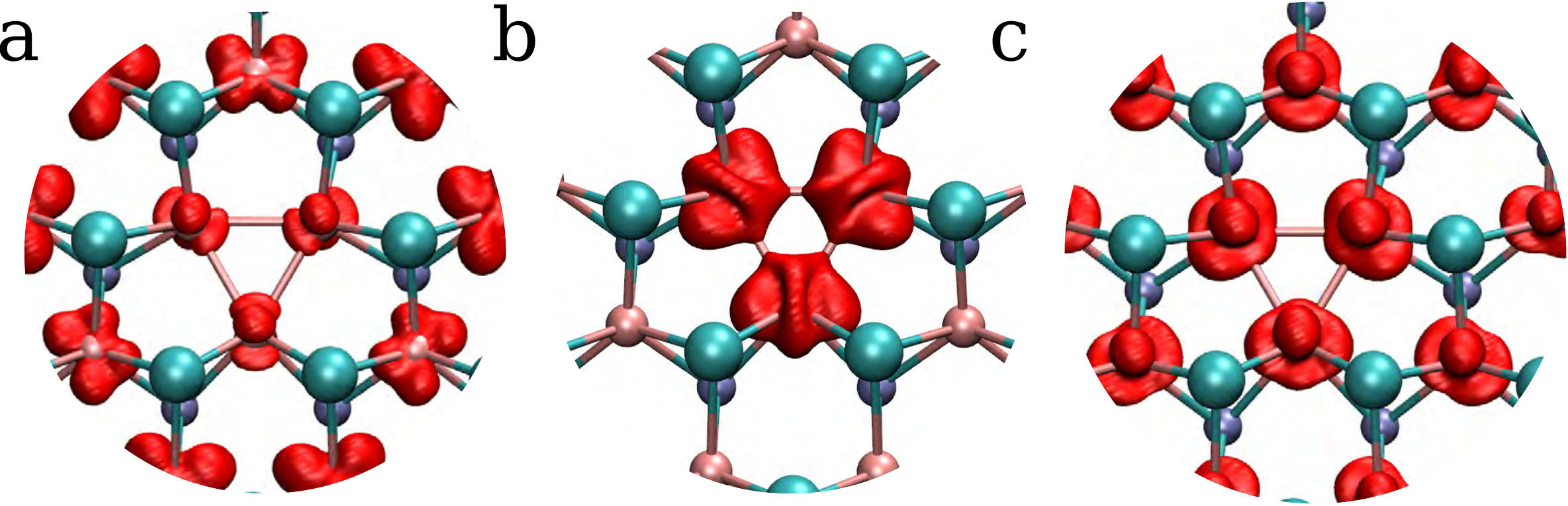}\\
		\includegraphics[scale=0.31]{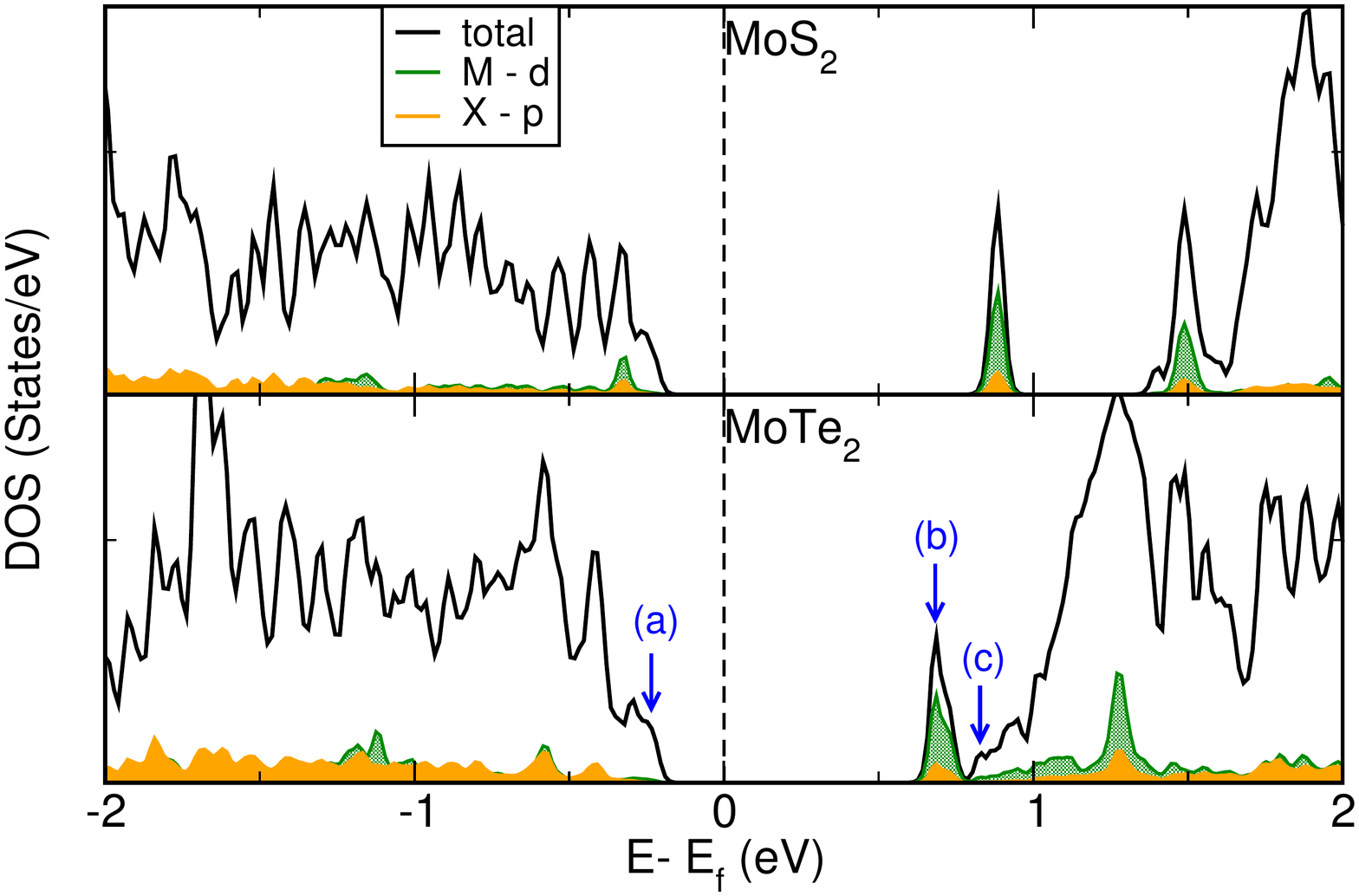}
		\caption{(Color online) Comparison of total density states and defect states for XX vacancy in MoS$_2$ and MoTe$_2$. Fig.~\ref{fig:xx-vac}(a)--(c) show partial charge density for defect peaks in the band gap region as noted in the DOS. See Fig.~\ref{fig:x-int} for other notations.}
		\label{fig:xx-vac}
	\end{center}
\end{figure}
\begin{figure}[!htbp]
	\begin{center}
		\includegraphics[scale=0.32]{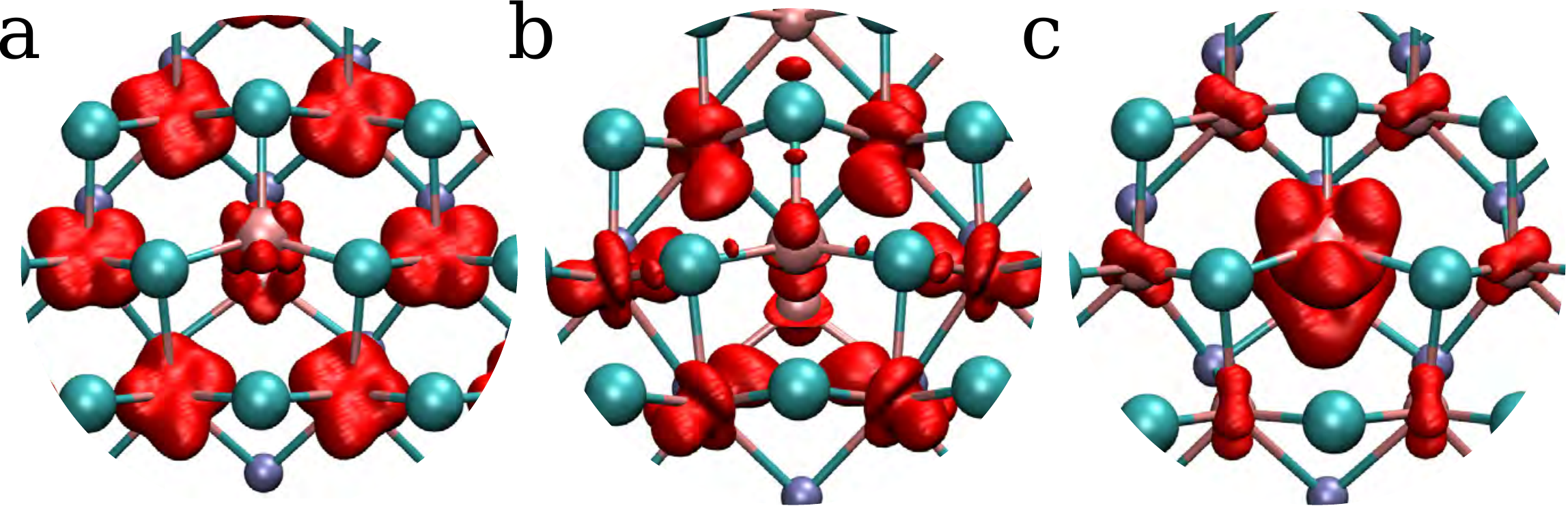}\\
		\includegraphics[scale=0.31]{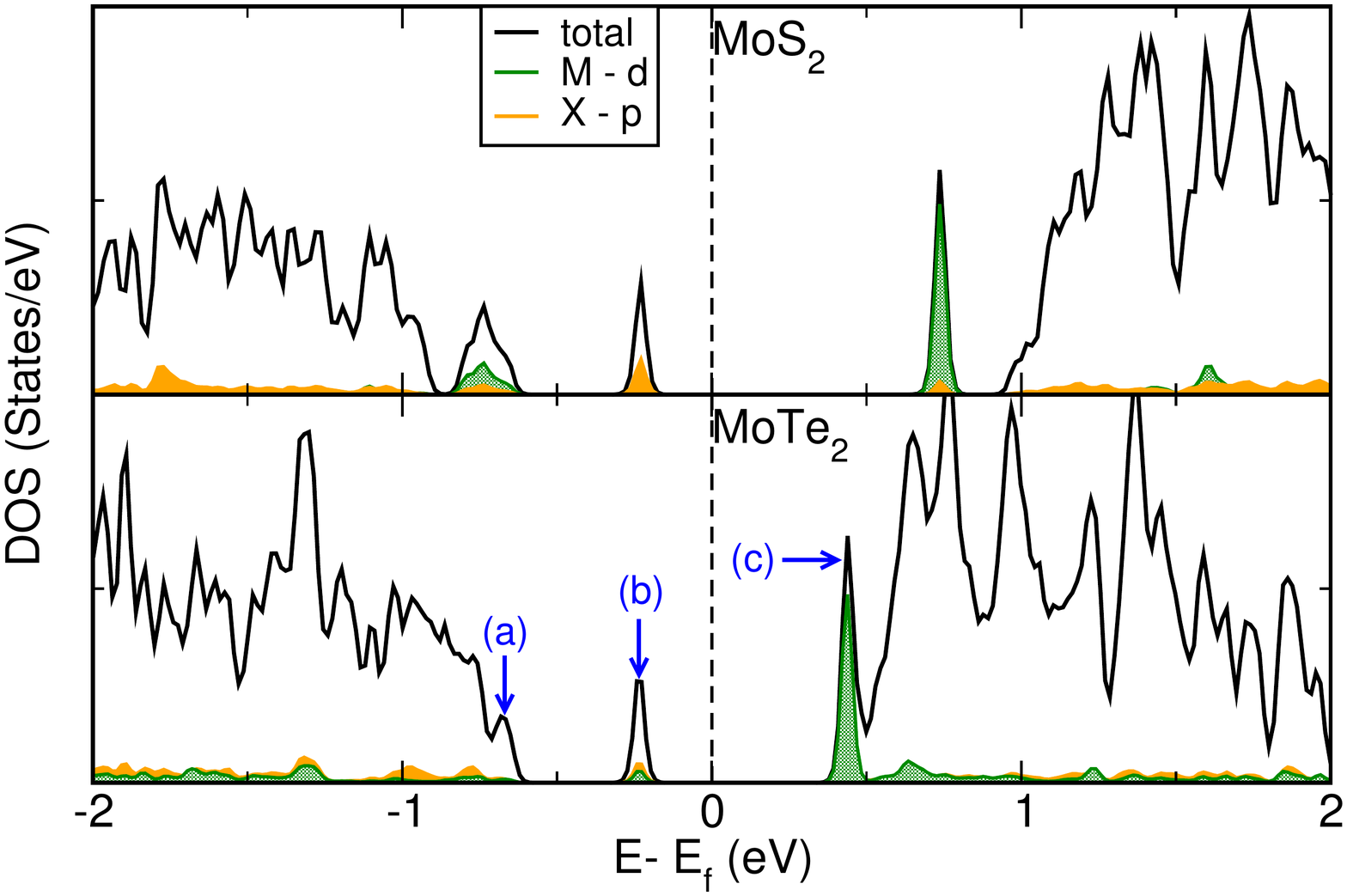}
		\caption{(Color online) Comparison of total density states and defect states for M interstitial in MoS$_2$ and MoTe$_2$. Fig.~\ref{fig:m-int}(a)-(c) show partial charge density for defect peaks in the band gap region as noted in the DOS. See Fig.~\ref{fig:x-int} for other notations.}
		\label{fig:m-int}
	\end{center}
\end{figure}
In this section, we will discuss the electronic properties of the defected MX$_2$ structures by analyzing total density of states (DOS), site and $m_l$  projected densities of states (PDOS) and partial charge densities. 

In Fig.~\ref{fig:dos-diag}, we have shown an energy level diagram for all the different defect states in all \ce{MX2} systems constructed from our calculated data. In each subplot, the columns (a)-(f) represent the (a) X--interstitial, (b) X--vacancy, (c) M--vacancy, (d) MX--vacancy, (e) XX--vacancy and (f) M--interstitial defect respectively. The orange long dashed line denote the position of valence band maxima (VBM) and conduction band minima (CBM) of the pristine \ce{MX2} systems. The short green solid and dashed lines denote the position of defect states for both occupied and unoccupied states respectively. From the analysis of energy level diagram, it can be seen that the defect states mainly appear in the gap region of the pristine system. Further analysis of DOS for all the defects in all the \ce{MX2} systems reveals that for a specific defect, the qualitative behavior is very much similar between \ce{MoS2}/\ce{WS2}, \ce{MoSe2}/\ce{WSe2} and \ce{MoTe2}/\ce{WTe2}. Furthermore, our analysis shows that the qualitative electronic properties due the presence of defects in \ce{MoS2}/\ce{MoSe2} are quite similar. Hence we have shown the DOS, PDOS analysis for only \ce{MoS2} and \ce{MoTe2} systems.  

The analysis of density of states for X interstitial defects for all the systems shows the same qualitative behavior. In Fig.~\ref{fig:x-int}(c), we have shown the total densities of states for two systems -- MoS$_2$ and MoTe$_2$. The general observation is that the defect states are found near the band edges. These states merge with both valence and conduction bands.
To find the orbital character of these defect states, we have carried out the analysis of site projected density of states (PDOS) and partial charge density analysis. Figs.~\ref{fig:x-int}(a) and \ref{fig:x-int}(b) show the partial charge density for valence band maximum (VBM) and conduction band minimum (CBM) respectively. Our analysis shows that the states near the VBM mainly arise from X$_i$ atom. These states are the doubly degenerately occupied $p_x$ and $p_y$ orbitals. There are also very small contributions of in-plane M -- $d_{xy}$ and $d_{x^2-y^2}$ states appearing in the VBM. The states near the CBM originate mainly from the M -- $d_{z^2}$ orbitals with some contribution from the unoccupied $p_z$ orbital of X$_i$ atom.

Fig.~\ref{fig:x-vac} shows density of states for X vacancy defects for MoS$_2$ and MoTe$_2$. For X vacancy defects, the features of density of states in the valence band side are identical for all the MX$_2$ systems. In these systems, the defect state appears at the valence band edge and inside the band gap towards the conduction band side.  The defect states can be further classified as singlet state (near valence band) and doubly degenerate state (the empty localized state in the gap) due to the trigonal symmetry of the X vacancy. The analysis of on-site projected DOS reveals that these states are formed from the combination of diagonal out of plane $d$ orbitals ($d_{xz}$ and $d_{yz}$) and in-plane $d$ orbitals ($d_{xy}$ and $d_{x^2-y^2}$) of the M atoms near the defected sites and very small amount of in-plane $p$ orbitals (mainly $p_x$) of second nearest neighbor X atoms. The orbital characters of VBM and CBM are respectively contributed by the $d_{x^2-y^2}$ and $d_{z^2}$ orbitals of the M atom.    

The analysis of PDOS and partial charge density shows that the defect states due to the M vacancy appear inside the gap region. The density of states for M vacancy in MoS$_2$ and MoTe$_2$ systems are shown in Fig.~\ref{fig:vac-m}. In this case, singlet and doublet defect states can be seen due to the trigonal and mirror symmetry of M vacancy and two layers of X atoms, respectively. The VBM contains both the $p_z$ orbital (significant contribution) of the six X atoms near the vacancy site and $d_{x^2-y^2}$ orbitals of the M atoms (see Fig.~\ref{fig:vac-m}(a)). The defect peak (marked as (b) in DOS) is a doublet state and originates mainly from the $d_{x^2-y^2}$ and $d_{z^2}$ orbitals of the six M atoms (second near neighbor from the vacancy site). The other defect peak (marked as (c) in the DOS) originates from the $p_z$ orbital of the six X atoms near the vacancy site along with very small $d_{xz}$ orbital contribution from the M atoms. The CBM mainly contains the  $d_{x^2-y^2}$ and $d_{z^2}$ orbitals of the M atoms.  
All these states are formed due to the local perturbation created in the geometry due to the absence of one M atom.

In Fig.~\ref{fig:mx-vac}, we have shown the MX--vacancy density of states comparison for \ce{MoS2} and \ce{MoTe2} as well as the partial charge density for the defect states, VBM and CBM. For MX vacancy, combined defect states due to M--vacancy and X--vacancy can be observed. As seen for the figure (see Fig.~\ref{fig:mx-vac}), the defect states appear mainly in the gap region. This defect has a strong influence on conduction band of \ce{MX2}. The states marked as (a) and (b) in the DOS) are mainly originating from the $d_{xy}$ and $d_{x^2-y^2}$ orbital of the M atoms near the vacancy site. A mix orbital character of $d_{xy}$, $d_{z^2}$ form the M atom and in-plane $p$ orbital from the dangling X atom can be observed in defect peak (c). However, the major contribution to the defect peak (d) is coming from the $p_z$ orbitals of the dangling X atom with a minor contribution from $d_{yz}$ orbital of M atom. The defect peak (e) is again a mixture of $d_{x^2-y^2}$, $d_{z^2}$ of M atom and $p_z$ of the X atom. However, the defect peak (f) is mainly of $d$ character with $d_{yz}$ and $d_{x^2-y^2}$ are being the dominant orbitals. The CBM mainly consists of $d_{z^2}$ orbital of M atoms.

For the XX double vacancy also the defect states can be seen in the gap region. Fig.~\ref{fig:xx-vac} shows the comparison of densities of states for MoS$_2$ and MoTe$_2$ and the partial charge densities of \ce{MoTe2} system for this defect. The analysis of partial charge density and PDOS shows that these defect states are the combined effect of two single X vacancies. Consequently, the $m_l$ characteristic features of the defect peaks are very much similar to the X--vacancy defects as discussed previously.

The density of states for Mo interstitial in \ce{MoS2} and \ce{MoTe2} and the partial charge density for the \ce{MoTe2} have been shown in Fig.~\ref{fig:m-int}. The general trend in the density of states for Mo interstitial defect in \ce{MX2} system is again very similar for all the systems. 
The defect levels are generated inside the band gap. The occupied states are one doublet and one singlet state and the unoccupied state is the doublet state. The contribution to the occupied defect states are mainly coming from the $d_{z^2}$ orbital of the interstitial M atom and the M atom attached with it. There is also a smaller contribution from the neighboring M atom's $d_{x^2-y^2}$ orbital. However, the contribution to the unoccupied defect state is manly from the defect site M atom's $d_{xy}$ and $d_{x^2-y^2}$ orbitals.

\subsection{Optical properties}
\label{subsec:optical}
\begin{figure}[hp]
	\begin{center}
		\includegraphics[scale=0.4]{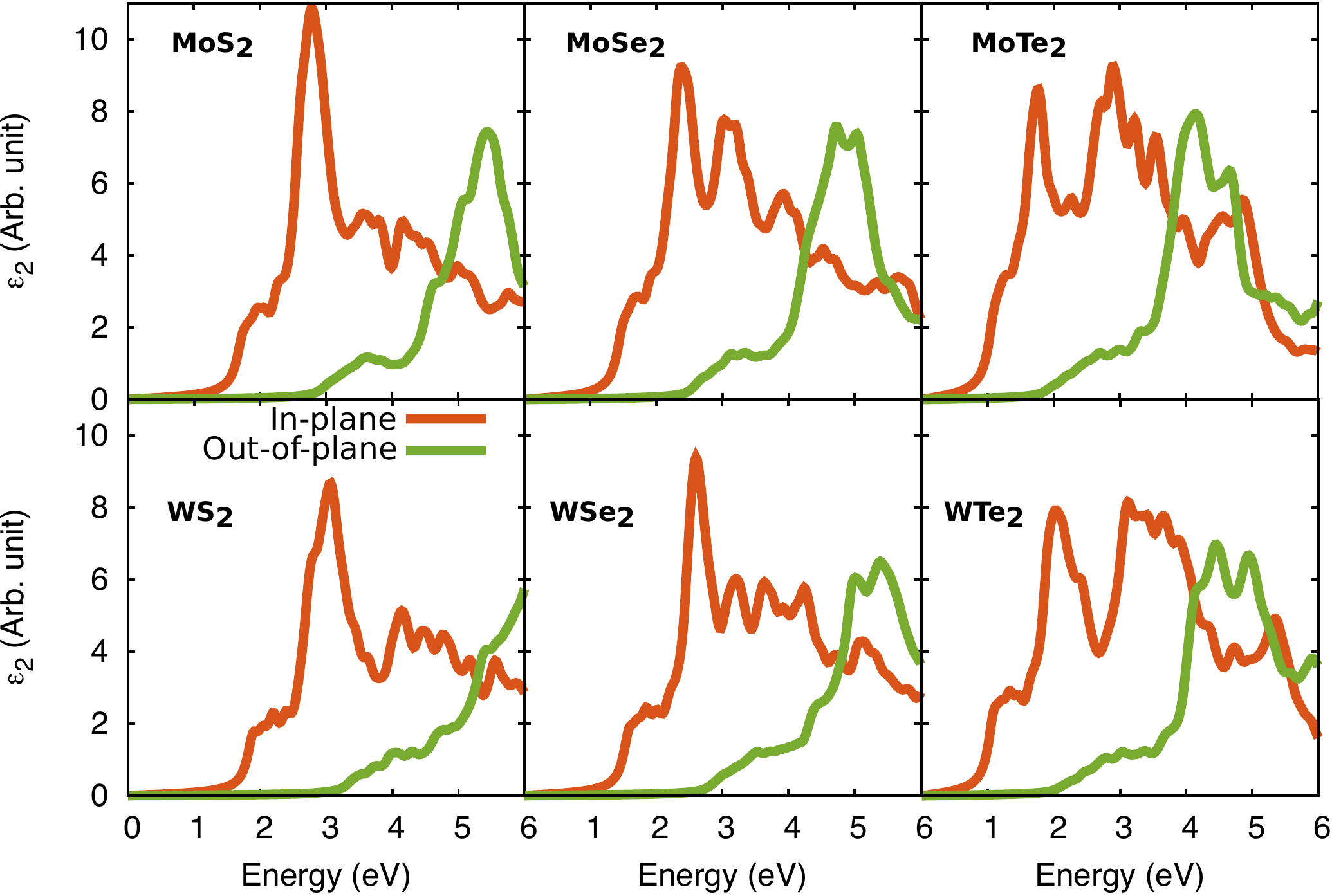}
		\caption{(Color online) Optical properties of pure \ce{MX2} system. Blue and orange lines denote In-plane and out-of-plane components of the imaginary parts of the dielectric functions ($\varepsilon_2$).}
		\label{fig:opti-pure}
	\end{center}
\end{figure}
\begin{figure}[htbp]
\begin{center}
\includegraphics[scale=0.38]{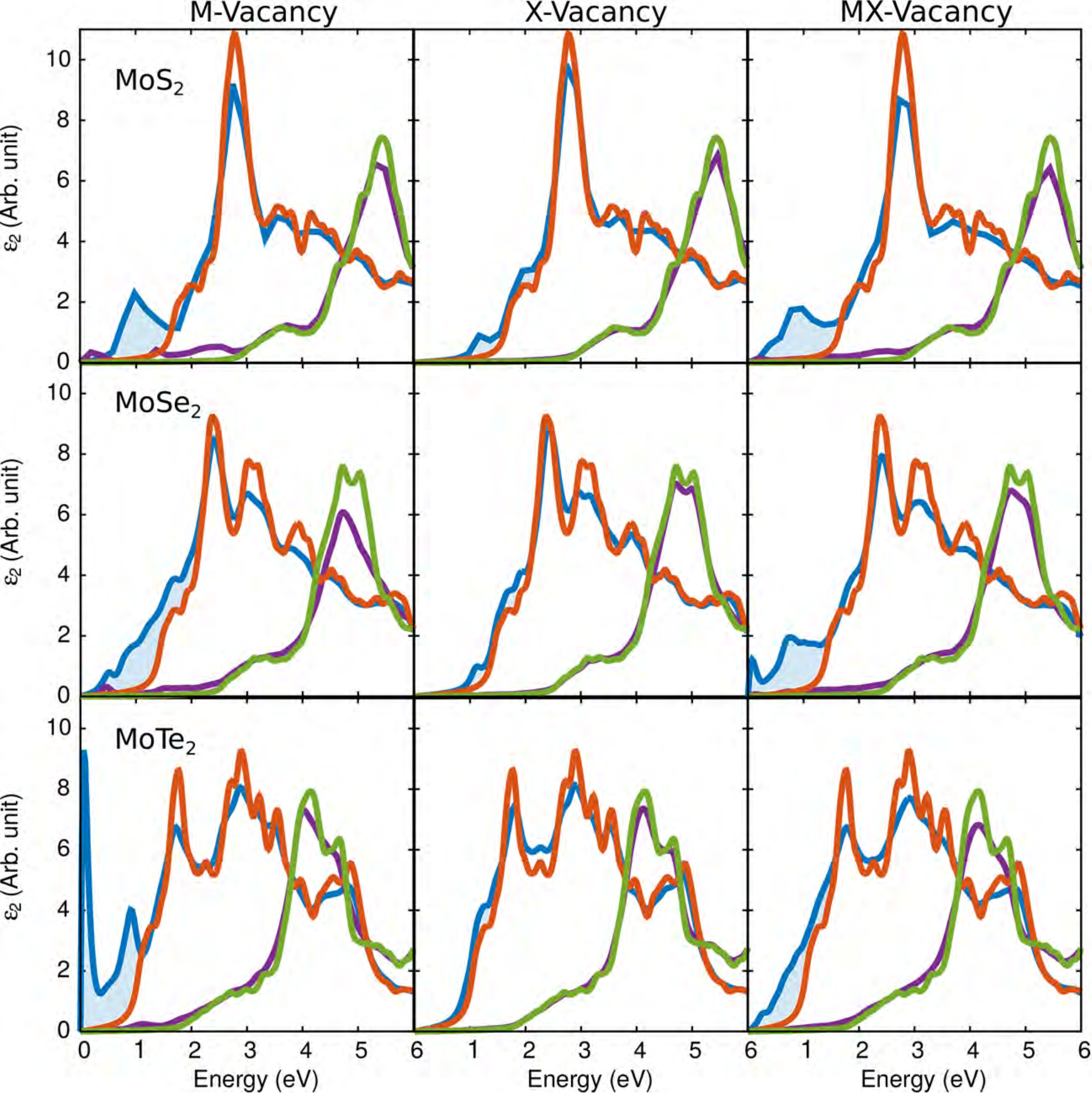}
\caption{(Color online) Comparison of imaginary parts of the dielectric functions ($\varepsilon_2$) between MoS$_2$, MoSe$_2$ and MoTe$_2$ for M--vacancy, X--vacancy and MX--vacancy. Blue and orange lines denote in-plane contribution of $\varepsilon_2$ for defected and pure systems respectively. Purple and green lines denote out-of-plane contribution of $\varepsilon_2$ for defected and pure systems respectively. The shaded region represents the prominent contributions of defect related peaks compare to the pristine system.}
\label{fig:opti-mo}
\end{center}
\end{figure}
\begin{figure}[htbp]
\begin{center}
\includegraphics[scale=0.38]{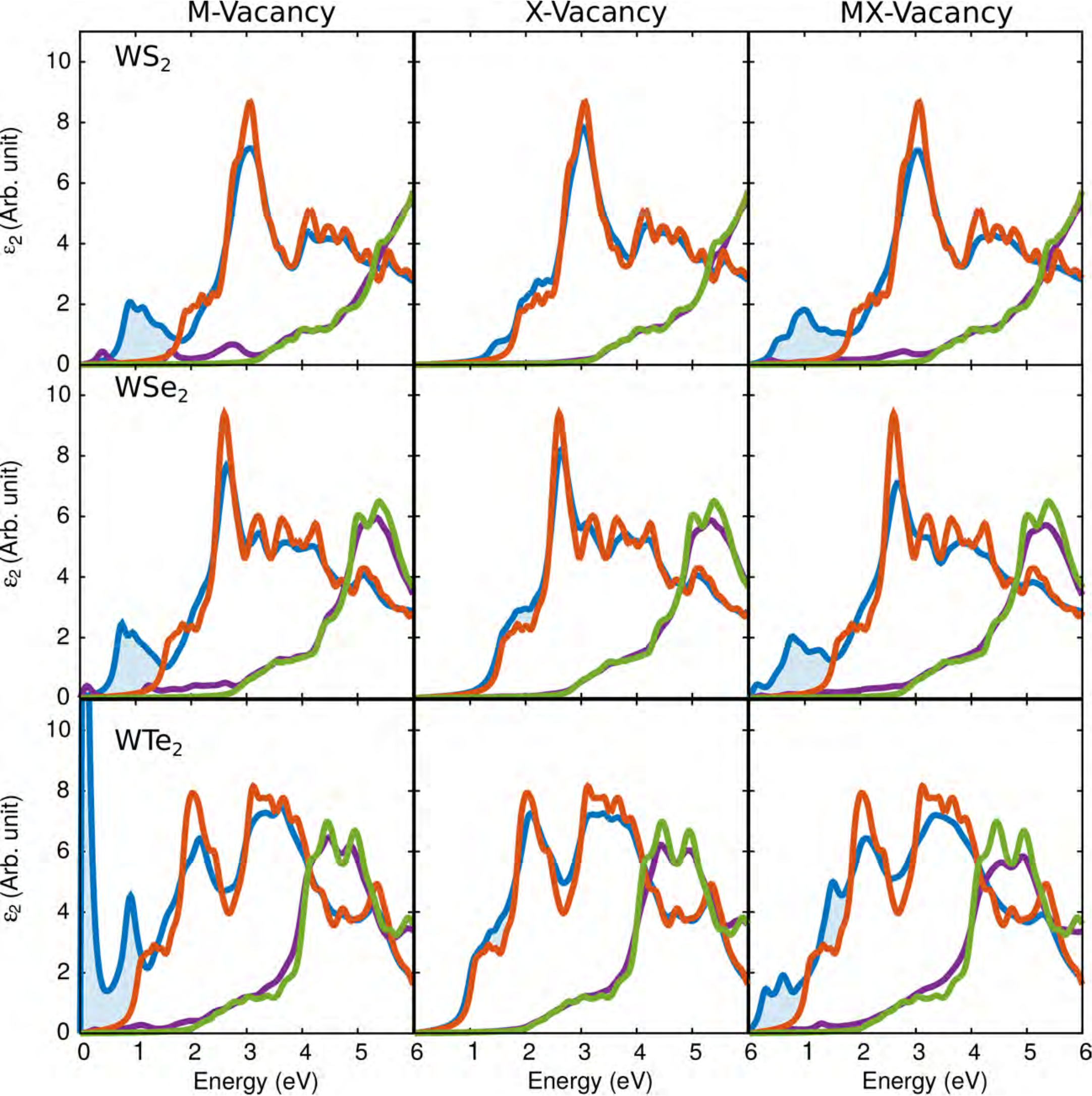}
\caption{(Color online) Comparison of optical properties between WS$_2$, WSe$_2$ and WTe$_2$ for M--vacancy, X--vacancy and MX--vacancy. See Fig.~\ref{fig:opti-mo} for other notations.}
\label{fig:opti-w}
\end{center}
\end{figure}

In addition to the geometric and electronic structures, we have also studied the influence of defects in the optical properties of defected \ce{MX2} by means of frequency dependent dielectric functions. In the following paragraphs we will briefly discuss the comparison of optical properties for the pristine system followed by the modification of optical properties due to the presence of defects. 
	
The comparison of optical properties for all the pristine \ce{MX2} systems have been shown in Fig.~\ref{fig:opti-pure}. We have plotted the imaginary parts of the dielectric functions with their in-plane ($xx$ and $yy$) and out of plane ($zz$) components as a function of energy. The $xx$ and $yy$ components have the same absorption spectra for all the systems and their absorption is stronger than the $zz$ component. From the figure, it can be seen that the general features of the optical spectra for same X elements are quite similar in nature with a step like peak followed by a dominant peak, valley and subsequent peaks. This step like peak is the characteristic of a two-dimensional system. From the analysis of the spectra, it can be seen that the prominent optical peak shifts towards lower energy and breaks into more than one peak as we move towards the heavier X element (see fig.~\ref{fig:opti-pure}). The step like peak also moves towards lower energy with heavier X element, signifying smaller band gaps. 

Introduction of defects in the pristine \ce{MX2} system can affect the optical properties also. From our calculations, we have found out that most notable change in the optical properties due to the presence of defect occurs for three following defects -- i) M--vacancy, ii) X--vacancy and iii) MX--vacancy. Therefore, in the following section we have discussed aforementioned defects only. 

In Fig.~\ref{fig:opti-mo} and Fig.~\ref{fig:opti-w},  we have shown the comparison of optical properties for M--, X-- and MX--vacancies for \ce{MoX2} and \ce{WX2} respectively.  As mentioned before,  here also the $xx$ and $yy$ components have the same absorption spectra for all the systems and their absorption is stronger than the $zz$ component.

For M--vacancy in \ce{MoS2}, the defect peaks appear just below and above the Fermi level (see fig.~\ref{fig:vac-m}), which have mainly $p$ and $d$ characters and are responsible for the electronic transition. This transition occurs at $\sim$ 1.0 eV. In \ce{MoSe2}, the peak is broader as there are two possible transitions possible with a very small energy interval. However for M--vacancy in \ce{MoTe2}, the defect peak appears at the Fermi energy (see fig.~\ref{fig:vac-m}) giving rise to a sharp optical transition at a very low energy (see fig.~\ref{fig:opti-mo}). 

For X--vacancy, the defect state appears very close to the CBM of pristine system (see fig.~\ref{fig:x-vac}). Thus the optical spectra with X--vacancy in different \ce{MoX2} systems do not change significantly from the pristine one.  In the case of MX--vacancy also a number of defect peaks appear near the Fermi energy for all \ce{MoX2} systems (See fig.~\ref{fig:mx-vac}) which give rise to the electronic transition at $\sim$ 0.5 eV with a relatively broader peak distribution than M--vacancy. 

The analysis of DOS for \ce{WX2} systems (figure not shown) show similar characteristics as of \ce{MoX2}. This leads to a very much similar optical properties as shown in Fig.~\ref{fig:opti-w} as compared to \ce{MoX2} (Fig.~\ref{fig:opti-mo}).

The calculated  dielectric functions elicit that the absorption spectrum appears in the visible region (3 eV--1.7 eV (400--700 nm)) for \ce{MX2} systems. It is well known that the material, which has absorption in the visible region is suitable for photocatalysis using sunlight. Therefore, \ce{MX2} having suitable defects can be used for the photocatalysis. Tongay {\etal}~\cite{tongay2013} in their study have investigated effects of anion vacancy on the photoluminescence of \ce{MoS2}, \ce{MoSe2} and \ce{WS2} where this vacancy is identified by a new peak appearing below the photoluminescence peak. Moreover, the intensity of the photoluminescence peak enhances with increasing defect density. As we discussed above, the absorption spectra of both M--vacancy and MX--vacancy defected \ce{MX2} have new peaks at $\sim$ 1.0 eV and $\sim$ 0.5 eV, respectively, which arise due to the defect states. This feature is very much similar to that observed with anion vacancy. Nowadays, it is possible to create point defects in 2D materials (as demonstrated in graphene) using ion-irradiation technique. One may speculate that the creation of suitable defects in \ce{MX2} may give rise to desired optical transitions suitable for light emitting diodes.

\section{Conclusions}
We have performed a systematic study of native defects, viz. X vacancy, X interstitial, M vacancy, M interstitial, MX vacancy and XX vacancy on a single layer of \ce{MX2} system (M = Mo, W; X = S, Se, Te) using first principles density functional theory. It has been found that under X--rich condition, X interstitial defect has the lowest formation energy for all the systems under investigation with equilibrium defect concentration $\sim \times 10^9$ cm$^{-2}$ at the growth temperature range of 1000-1200K. For M--rich environment, X vacancy has the lowest formation energy except \ce{MTe2} systems, where X interstitial defect has the lowest formation energy. Metal atom and double vacancy related defects are quite high in formation energy thus showing almost zero equilibrium defect concentration at the studied growth temperatures. Our calculations reveal the position of defect states in the band gap along with the orbitals contributing to those states. Finally, our calculated optical properties indicate prominent signatures in the in-plane component of the imaginary part of the dielectric functions. One may speculate that suitably designed defected \ce{MX2} systems can be promising materials for light emitting devices. 

\section*{Acknowledgement} 
OE acknowledges KAW foundation for financial support. In addition, BS acknowledges Carl Tryggers Stiftelse, Swedish Research Council and KOF initiative of Uppsala University for financial support. SH also acknowledges valuable discussions with Carmine Autieri. We are grateful to NSC under Swedish National Infrastructure for Computing (SNIC) and the PRACE project resource Cy-Tera supercomputer based in Cyprus at the Computation-based Science and Technology Research Center (CaSToRC) and Zeus supercomputer based in Poland at the academic computer center Cyfronet. Structural figures were generated using VMD~\cite{vmd}.


\bibliography{biblio}

\end{document}